# Giant periodic pseudo-magnetic fields in strained kagome magnet FeSn epitaxial films on SrTiO$_3$(111) substrate

Huimin Zhang[1,2], Michael Weinert[3], and Lian Li[1*]

[1]Department of Physics and Astronomy, West Virginia University, Morgantown, WV 26506, USA

[2]State Key Laboratory of Structural Analysis, Optimization and CAE Software for Industrial Equipment, Dalian University of Technology, Dalian, 116024, China

[3]Department of Physics, University of Wisconsin, Milwaukee, WI 53201, USA

[*]Correspondence to: lian.li@mail.wvu.edu

**Abstract:**

Quantum materials, particularly Dirac materials with linearly dispersing bands, can be effectively tuned by strain-induced lattice distortions leading to a pseudo-magnetic field that strongly modulates their electronic properties. Here, we grow kagome magnet FeSn films, consisting of alternatingly stacked Sn$_2$ honeycomb (stanene) and Fe$_3$Sn kagome layers, on SrTiO$_3$(111) substrates by molecular beam epitaxy. Using scanning tunneling microscopy/spectroscopy, we show that the Sn honeycomb layer can be periodically deformed by epitaxial strain for film thickness below 10 nm, resulting in differential conductance peaks consistent with Landau levels generated by a pseudo-magnetic field greater than 1000 T. Our findings demonstrate the feasibility of strain engineering the electronic properties of topological magnets at the nanoscale.

Dirac materials are characterized by linearly dispersive energy bands, hosting massless Dirac fermions[1]. When subjected to strain, which introduces position-dependent perturbations, electronic properties can be significantly impacted[2]. For graphene, a prototypical Dirac material, structural distortions due to strain can modify the hopping energies between π orbitals in different sublattices, which causes the Dirac points to shift to opposite directions analogous to an applied out-of-plane magnetic field, giving rise to a pseudo-quantum Hall effect[3,4]. Different from the real magnetic field, however, time-reversal symmetry is preserved with a pseudo-magnetic field[2]. As a result, the vector potential generated by the strain has opposite signs at the two K valleys, thus leading also to a valley Hall effect[5,6]. Strain in graphene has been induced by either geometrical confinements such as in nanobubbles[7] and ripples[6,8], nanoscale strain engineering[9–12], or lattice mismatch in heterostructures such as graphene/(NbSe$_2$, BN)[13] and graphene/black-phosphorus[14]. Pseudo-magnetic field up to 800 T has been reported[12]. Beyond graphene, similar phenomena have also been predicted for other Dirac materials, including Dirac and Weyl semimetals[15–19], Weyl superconductors[16,20–23]. However, experimental observations are limited to a recent report of strain-induced Landau levels on the surface of cleaved Weyl semimetal Re-doped MoTe$_2$[24], where a moderate 3 T field is reported. Furthermore, the strain-induced pseudo-magnetic field is similarly calculated for kagome lattice, characterized by a two-dimensional hexagonal network of corner-sharing triangles, leading to linearly dispersing Dirac states at the K point and flat band through the rest of the BZ[25]. There is no experimental report of pseudo-magnetic field in kagome materials.

Here, we provide strong evidence for strain-induced pseudo-magnetic field over 1000 T in epitaxial FeSn films grown on the SrTiO$_3$(111) (STO) substrate by molecular beam epitaxy (MBE). The model kagome magnet FeSn consists of alternatingly stacked 2D kagome Fe$_3$Sn (K) and honeycomb Sn$_2$ (S) layers (Figure 1a)[26–28]. The S-layer, or stanene, has a quasi-2D electronic structure[26,29], similar to graphene. However, the stanene's Sn-Sn bond is weaker than the C-C bonding in graphene due to a larger bond length, thus facilitating a greater degree of its distortion or deformation. Using scanning tunneling microscopy/spectroscopy (STM/S), we show that the Sn honeycomb can be significantly distorted by epitaxial strain, leading to periodic stripe modulations with a periodicity of $l$ = 2.0 nm, ~3.8 $a_{FeSn}$. Such modulations also result in differential conductance peaks consistent with pseudo-Landau levels generated by pseudo-magnetic fields of over 1000 T. Our results demonstrate the feasibility of strain engineering topological magnets at the nanoscale.

**MBE growth of FeSn films on SrTiO$_3$(111) substrates.** The FeSn films are grown on STO substrates, which are thermally treated *in-situ* to obtain a flat-surface morphology with a (4 × 4) reconstruction[30] (STM images of the annealed SrTiO$_3$(111) are provided in Figure S1, Supporting Information). Due to an in-plane lattice mismatch between FeSn ($a_{FeSn}$ = 5.30 Å)[31] and SrTiO$_3$(111) ($a_{STO(111)}$ = 5.52 Å)[30], a tensile strain $\varepsilon$ = 3.99% is expected in epitaxial FeSn films. The growth follows the Volmer-Weber mode, i.e., island growth, at $T_{sub}$ between 480 and 530 °C, characterized by three-dimensional flat-top islands as revealed by topographic STM image in Figure 1**b** and line-profiles shown in Supporting Information Figure S2. The growth of the FeSn phase is confirmed by x-ray diffraction (XRD), which shows diffraction peaks of FeSn (002) and (021) planes (Supporting Information, Figure S3). As FeSn consists of vertically stacked Fe$_3$Sn kagome and Sn$_2$ honeycomb layers, there are two possible interfaces with the SrTiO$_3$ substrate. An earlier study indicates a complex interface with Fe$_3$Sn kagome layer on a Ti-rich termination layer of the SrTiO$_3$ substrate[32]. For FeSn films studied here, both surface terminations are observed (STM images of mixed termination and K-layer are shown in Supporting Information, Figures S4-5) with the Sn-termination the most common, likely due to Sn-rich growth conditions (Supporting Information Note 1). In the region outlined by the white square, atomic resolution imaging reveals a perfect honeycomb structure (Figure 1**c**), which is also independent of the bias voltage (Supporting Information, Figure S6).

The majority of the islands, however, exhibits periodic stripe modulations with an average periodicity of *l* = 2.0 nm, ~3.77 $a_{FeSn}$. A few examples are marked by black, cyan, and yellow arrows in Figure 1**b**. The stripes are distributed along three directions, ~150° apart (Figure S7, Supporting Information). As shown in Figure S2, the stripes are commonly observed on islands less than 10 nm thick, which are likely more strained. For films grown at higher temperature of 530 °C, the island density is slightly reduced with larger lateral size and increased height, without stripe modulations on the surface. The fact that the thicker films are still under strain is further confirmed by *ex-situ* XRD measurement, where the FeSn (002) and (021) peaks are slightly shifted to higher values, indicating smaller lattice constant in the *c*-direction, which is typically accompanied by an expansion of the in-plane constant. However, strain in these thicker films is likely not enough to distort the Sn or $Fe_3Sn$ layer.

***Strain-induced strong distortion of the Sn honeycomb lattice.*** Atomic resolution imaging further reveals that the stripes arise from distortion of the Sn honeycomb lattice. As marked by the ball-and-stick model in Figure 1**d**, the building block is a group of four slightly distorted honeycombs periodically shifted by one-and-half unit bond length along the ***x***-direction, forming a stripe at $\theta$ = 14.3° with respect to the ***y***-direction. Between the stripes along the ***x***-direction, there are three strongly distorted honeycombs, where the middle unit exhibits the largest distortion (highlighted in white). The angles between the bonds (as marked) deviate significantly from the 120° of a perfect honeycomb.

The stripe formation is further confirmed by fast Fourier Transformation (FFT) analysis of STM images. An example is shown in Figure 2**a**, where the periodic stripe modulation of the Sn honeycomb is apparent. In addition to the Bragg peaks $Q_1$, $Q_2$, and $Q_3$ characteristic of the honeycomb lattice, the 4$^{th}$ feature $Q_4$ is also observed (Figure 2**b**), as the result of the stripe pattern, which is further confirmed by the selective reverse-FFT image (Figure 2**c**). The stripe forms a 13.8° angle with respect to $Q_1$, consistent with that calculated from the real space image (Figure 1**d**). Furthermore, the Bragg peaks $Q_1$, $Q_2$, and $Q_3$ exhibit different lengths compared to that of $Q$ for a perfect honeycomb lattice (Figures 2**d**&**e**). For a quantitative comparison, calibration of the lattice distortion caused by the nonlinearity of the STM scanner is carried out first (Supporting Information Figures S8-9). The results show that while $Q$ corresponds to a lattice constant of 0.533 nm, the Bragg peaks $Q_1$, $Q_2$, and $Q_3$ correspond to lattice constants of 0.526, 0.544, and 0.518 nm, respectively. This leads to an average 1.26% and 2.85% compressive strain along the $q_1$ and $q_3$ directions, respectively, and a 2.16% tensile strain along the $q_2$ direction. The analysis is consistent with that observed in real space. For example, the 2.16% tensile strain along $q_2$ direction is normal to the orientation of stripes (Figures 2**a**&**b**). Overall, the perfect honeycomb is likely deformed by the epitaxial strain, leading to different lattice vectors $a_1$ and $a_2$ (Figures 2**f**&**g**). These strain-induced distortions thus break the $C_3$ symmetry of the stanene layer (albeit one direction is more pronounced), satisfying the conditions to generate pseudo-magnetic fields[3,33].

As STM imaging is a convolution of structural and electronic contributions, a natural question arises whether the distortion of the honeycomb lattice is structural or electronic. The stripe modulations indeed exhibit bias-dependence, with the distortion the most obvious at bias voltages closer to the Fermi level (Analysis of the bias-dependent stripe modulations is provided in Supporting Information Figures S10-12). While the relative amplitude varies from 8 to 20 pm depending on the imaging bias voltage, the feature associated with stripe modulations is always observed in the FFT patterns at all energies. We can also rule out the possibility of tip artifacts based on imaging of stripes across different domains (Supporting Information Figure S7). In addition, the stripe modulations are only observed on the Sn layer and not the $Fe_3Sn$ kagome layer. An example is shown in Supporting Information Figure S4 for an island with mixed S- and K-termination, where only the S-termination is distorted while a close-packed structure is intact on the K-termination. This is likely due to the weaker bonding in the S-layer facilitating greater lattice deformation. Overall, these observations indicate that the stripe modulations are primarily structural.

***Periodic modulations of FeSn electronic structure.*** For the perfect honeycomb lattice (Figure 3**a**), spatially uniform dI/dV spectra are observed (Figure 3**b**), characterized by three features: a gap near Fermi level, one dip at -0.15 (cyan arrow and dashed line), and another at -0.36 eV (black arrow and dashed line). The later feature, with an averaged energy position of -0.36 ± 0.01 eV, is attributed to the Dirac point $E_D$, similar to the -0.4 eV reported in recent photoemission studies on the Sn-terminated surface of cleaved bulk FeSn[26,27]. The slight shift of the Dirac point in FeSn/STO films relative to the bulk value is likely due to charge doping from the $SrTiO_3$(111) substrate. While the origins of the gap near the Fermi level and dip at -0.15 eV are unknown, all three features are modulated by the formation of stripes as revealed by spatially resolved differential dI/dV spectroscopy discussed below.

Figure 3**d** shows a series of dI/dV spectra taken at sites 1-15 marked in Figure 3**c**. Spectra taken on the stripe (weakly distorted honeycomb) and between stripes (strongly distorted honeycomb) are further highlighted in Figure 3**e**. Compared to the perfect honeycomb lattice, a new peak at ~0.4 eV above $E_F$ (purple arrow in Figure 3**e**) appears for the weakly distorted honeycomb (on stripe). Below $E_F$, a pronounced peak around $E_D$ = -0.36 eV emerges for the strongly distorted honeycombs (between the stripes) (red arrow in Figure 3**e**). The same feature is consistently observed at all seven characteristic sites within the strongly distorted honeycomb (Figure 3**f**), which suggests that the electronic modulation follows the structural modulation, extending beyond the individual unit cell of the honeycomb, and consistent with the stripe distribution. For dI/dV spectra taken on stripes, slight spatial variation is observed (additional dI/dV spectra provided in Figure S13). The dI/dV spectra are also independent of the tunneling current *I*. The lineshape of the dI/dV spectra remains similar as current is varied by almost two orders of magnitude from 0.1 to 7.0 nA, with only variations in spectral intensity (Supporting Information, Figure S14). Note that the gap $\Delta \sim 7.9$ meV at $E_F$ is observed on the Sn-terminated surface without (Figure 3**b**) or with stripe modulations (Figures 3**d&f**). Therefore, it is unlikely a charge density wave gap. This is further confirmed by systematic analysis of energy-dependent dI/dV maps where the intensity of the stripe modulation is uncorrelated to this gap (Supporting Information, Figure S15). The origin of the gap is unknown at the moment, but may be due to electron-electron interactions, as that observed in 1T'-WTe$_2$[34].

***Periodic strain-induced pseudo-magnetic field.*** The periodic stripe modulations also lead to regularly modulated electronic properties. Figure 4**a** shows a region with ten periodic stripe modulations, where n' and n label the on- or between stripe sites. Figure 4**b** is a false color plot of a series of dI/dV spectra taken along the white arrow in Figure 4**a** (dI/dV spectra are shown in Figure S16, Supporting Information). Periodic variations of the differential conductance are clearly resolved, consistent with the topographic stripe modulations in Figure 4**a**. To quantitatively determine the modification in electronic structures induced by strain, differential spectra on- and between stripe sites (n-n') are obtained. An example is shown in Figure 4**c**, which reveals three peaks marked by black arrows and dashed lines (lower panel). Similar analysis was done for the rest of the spectra (Figure 4**d**), where the same peaks are also observed as marked by black dashed lines (Figure S17, Supporting Information).

The appearance of these conductance peaks could be due to quantum well states from spatial confinement[35], given that the islands are mostly less than 100 nm in diameter (c.f. Figure 1**b**). However, periodic modulations of the dI/dV spectra were only observed on islands that are less than 10 nm in height, but not on taller islands of comparable size (Supporting Information Figure S2). On the later type of islands, bias-dependent imaging also did not reveal any modulations (c.f. Supporting Information Figure S6). Another possible mechanism is defect-induced states. The most commonly observed defects are Sn divacancy and substitutional defects, as shown in Supporting Information Figure S18, both of which do induce bound states appearing as peaks in dI/dV spectra. However, their line shapes, particularly the single peak position at -86 meV and -61 meV, respectively, are intrinsically different from those caused by the stripe modulations.

After ruling out these possible mechanisms, we attribute the conductance peaks to quantized Landau levels (LLs) originating from the strain-induced pseudo-magnetic field. Such field is a general response of materials with linear energy dispersion to strain, which generates axial vector and scalar potentials[4]. At the level of tight binding, the vector potential will modify the massless Dirac Hamiltonian as follows[4]:

$$-i\hbar v_F \vec{\sigma}(\nabla - i\vec{A}) \qquad (1)$$

similar to that from a real magnetic field. This will lead to LLs that are expected to follow [36,37]:

$$E_N = E_D + sgn(N) v_F \sqrt{2e\hbar |N| B_{eff}} \qquad (2)$$

Here, $E_D$ is the energy of Dirac point, $v_F$ is the Fermi velocity, $e$ is the electron charge, $\hbar$ is the reduced Planck constant, and $B_{eff}$ is the pseudo-magnetic field. The integer $N$ represents an electron-like ($N > 0$) or a hole-like ($N < 0$) LL index.

For the dI/dV spectra shown in Figure 4**d**, the most pronounced peak at $E$ = -0.36 eV coincides with the Dirac point $E_D$, and is assigned to the 0$^{th}$ Landau level (labeled $N$ = 0). Starting from this charge neutrality point $N$ = 0 LL, the neighboring two LLs are labeled as $N$ = ±1. The plot of $E_N -$ $E_D$ as a function of $sgn(N)\sqrt{|N|}$ is shown in Figure 4**e**, which follows the unique square root dependent sequence, confirming the expected scaling behavior for LLs. The linear fit yields a slope *k* = 0.2182 ± 0.006. To calculate the value of the pseudo-magnetic field, Fermi velocity is needed. Since Fermi velocity for FeSn thin films is not available, we adopt the values for bulk FeSn of $v_F = (1.7 \pm 0.2) \times 10^5 \, ms^{-1}$ [26], which yields an effective pseudo-magnetic field of $B_{eff}$ = 1251 ± 363 T (details in Supporting Information Note 2). Note that a recent work[32] shows that $v_F$ can depend strongly on the inter-layer coupling between the kagome and the stanene layers in in FeSn thin films, where a significant decrease of $v_F$ can be expected at the bilayer limit. This suggests a low limit for our estimated effective pseudo-magnetic field. On the other hand, a lower limit can be obtained by using the Fermi velocity of free-standing stanene[38] (see Supporting Information Note 2 for details).

While pseudo-magnetic fields up to 800 T had been previously reported in graphene in nanobubbles or nanocrystals[7,13], our results show a periodic pseudo-magnetic field in excess 1000 T in strained kagome magnet FeSn/SrTiO$_3$(111) films grown at temperatures below 500 °C. Moreover, unlike the localized nature of the geometrical deformation in most graphene nanostructures, the pseudo-magnetic fields found here distribute periodically across the surface. This can be clearly seen in the differential conductance map *g*(**r**, -300 meV) near the 0[th] LL with periodic higher (lower) contrast for the regions between the stripes (on the stripes) (c.f. Supporting Information Note 3 and Fig. S19). Similar behavior is found for multiple samples as shown in Supporting Information Figures S20-21, where the induced pseudo-magnetic field is also independent on the stripe orientations.

The large pseudo-magnetic field can be attributed to several factors. First, the Sn-Sn bond length in single layer Sn honeycomb lattice, or stanene, is predicted to be 2.87 Å, the longest amongst all group IV honeycomb lattices, and two times larger than that of the C-C bond length in graphene (1.42 Å)[39]. The longer Sn-Sn bond length means a weaker π-π bond, which can facilitate the formation of a buckled structure with a mixed character of sp$^2$ and sp$^3$ orbital hybridization[39]. This makes the synthesis of freestanding stanene challenging, and most films synthesized to date exhibit a close-packed structure due to the vertical displacement of Sn atoms[40]. On cleaved surfaces of kagome materials, the formation of a perfect Sn honeycomb lattice depends on the stacking order of the Sn and kagome layers. For CoSn consisting of alternating Sn and $Co_3Sn$ layers[41], a perfect honeycomb structure was seen, while a buckled honeycomb structure was observed on $Fe_3Sn_2$ where the Sn layer is separated by two $Fe_3Sn$ kagome layers[42,43]. Here, our comprehensive STM/S studies of multiple FeSn films epitaxially grown on $SrTiO_3$(111) substrates have revealed a perfect Sn honeycomb lattice for those grown at temperature $T$ = 530 °C. At lower temperatures, the stripes are commonly observed on the Sn-terminated FeSn islands with thickness below 10 nm, attributed to epitaxial strain resulting in a distorted honeycomb lattice. These findings indicate that in addition to uniform buckling, the Sn honeycomb can also form long-range deformation for strain-relief, highlighting the significant tunability of epitaxial kagome thin films. Buckling of the honeycomb structure has also been found to be critical in the 2D magnetism of silicene and germaneness in Zintl phase compounds[44,45], indicating that strain engineering can be a fertile ground for tuning these types of materials for spintronic applications.

In summary, we have grown thin films of kagome magnet FeSn on the $SrTiO_3$(111) substrates by MBE and observed strain-induced periodic modulations of the Sn honeycomb lattice for film thickness less than 10 nm. Such modulations lead to periodic differential conductance peaks consistent with Landau levels generated by pseudo-magnetic fields greater than 1000 T. Our findings demonstrate a viable path towards strain engineering electronic properties of kagome materials.

**METHODS**

**Sample preparation.** The FeSn films were grown by MBE on Nb-doped (0.05 wt.%) $SrTiO_3$(111) substrates, which were were first degassed at 600 °C for 3 hours, and then followed by annealing at 950 °C for 1 hour to obtain a flat surface with step-terrace morphology. During the MBE growth, high purity Fe (99.995%) and Sn (99.9999%) were evaporated from Knudson cells setting at 1150 °C and 805 °C on the $SrTiO_3$(111) substrate at temperatures between 480 to 530 °C. The Fe/Sn ratio is estimated to be 1:2.7 during growth.

**LT-STM/S characterization.** The STM/S measurements were carried out at 4.5 K in a Unisoku ultrahigh vacuum low-temperature STM system. Polycrystalline PtIr tips were used and tested on Ag/Si(111) films before the STM/S measurements. The dI/dV tunneling spectra were acquired using a standard lock-in technique with a small bias modulation $V_{mod}$ = 20 mV at 732 Hz.

**ASSOCIATED CONTENT**

**Supporting Information**

The Supporting Information is available free of charge at

MBE growth of FeSn films on SrTiO$_3$(111) substrate; Calculations of the pseudo-magnetic field; Modulation of electronic properties by stripes revealed by dI/dV maps; STM imaging of the annealed SrTiO$_3$(111) substrate; Initial stages of FeSn growth on SrTiO$_3$(111) substrates; X-ray diffraction (XRD) data of a 20 nm FeSn/SrTiO$_3$(111) film; STM imaging of the mixed Sn termination (S layer) and Fe$_3$Sn kagome termination (K layer) in FeSn/STO(111) films; Bias-dependent STM imaging of the kagome layer on the K-terminated FeSn/SrTiO$_3$(111) film; Bias-dependent STM imaging of the Sn honeycomb on the Sn terminated FeSn/SrTiO$_3$(111) film; Stripe modulations on the Sn-terminated FeSn/SrTiO$_3$(111) films; FFT of perfect and deformed honeycomb lattices; Calibration of lattice distortion caused by nonlinearity of the STM scanner; Bias-dependent STM imaging of distorted honeycomb lattice; Bias-dependence of stripe modulations; FFT of bias-dependent STM images; Site-dependent dI/dV spectra within one honeycomb lattice; Setpoint-dependent of dI/dV spectra on stripe (n') and between stripes (n); FFT analysis of the stripe modulations; Periodic stripe modulations; Subtraction of 10 sets of dI/dV spectra to reveal the Landau levels; Bound states associated with two types of defects on the Sn-terminated FeSn/STO(111) film; Spatial-resolved dI/dV maps; Another example of Landau levels by pseudo-magnetic field; Laudau levels independent on the stripe orientations (PDF)

**AUTHOR INFORMATION**

**Corresponding Author**

Lian Li − Department of Physics and Astronomy, West Virginia University, Morgantown, West Virginia 26506, United States

Phone: (+1) 304-293-4270; E-mail: lian.li@mail.wvu.edu

Authors

**Huimin Zhang** − Department of Physics and Astronomy, West Virginia University, Morgantown, West Virginia 26506, United States

State Key Laboratory of Structural Analysis for Industrial Equipment, Dalian University of Technology, Dalian, 116024, China

**Michael Weinert** − Department of Physics, University of Wisconsin, Milwaukee, WI 53201, USA

Author contributions

L.L. and H.Z. conceived and organized the study. H.Z. performed the MBE growth and STM/S measurements. All authors analyzed the data, and H.Z. and L.L. wrote the paper.

Notes

The authors declare no competing financial interest.

ACKNOWLEDGMENTS

Research supported by the U.S. Department of Energy, Office of Basic Energy Sciences, Division of Materials Sciences and Engineering under Award No. DE-SC0017632, the U.S. National Science Foundation under Grant No. EFMA-1741673, and the Office of Naval Research under Award No. N00173-22-1-G001.

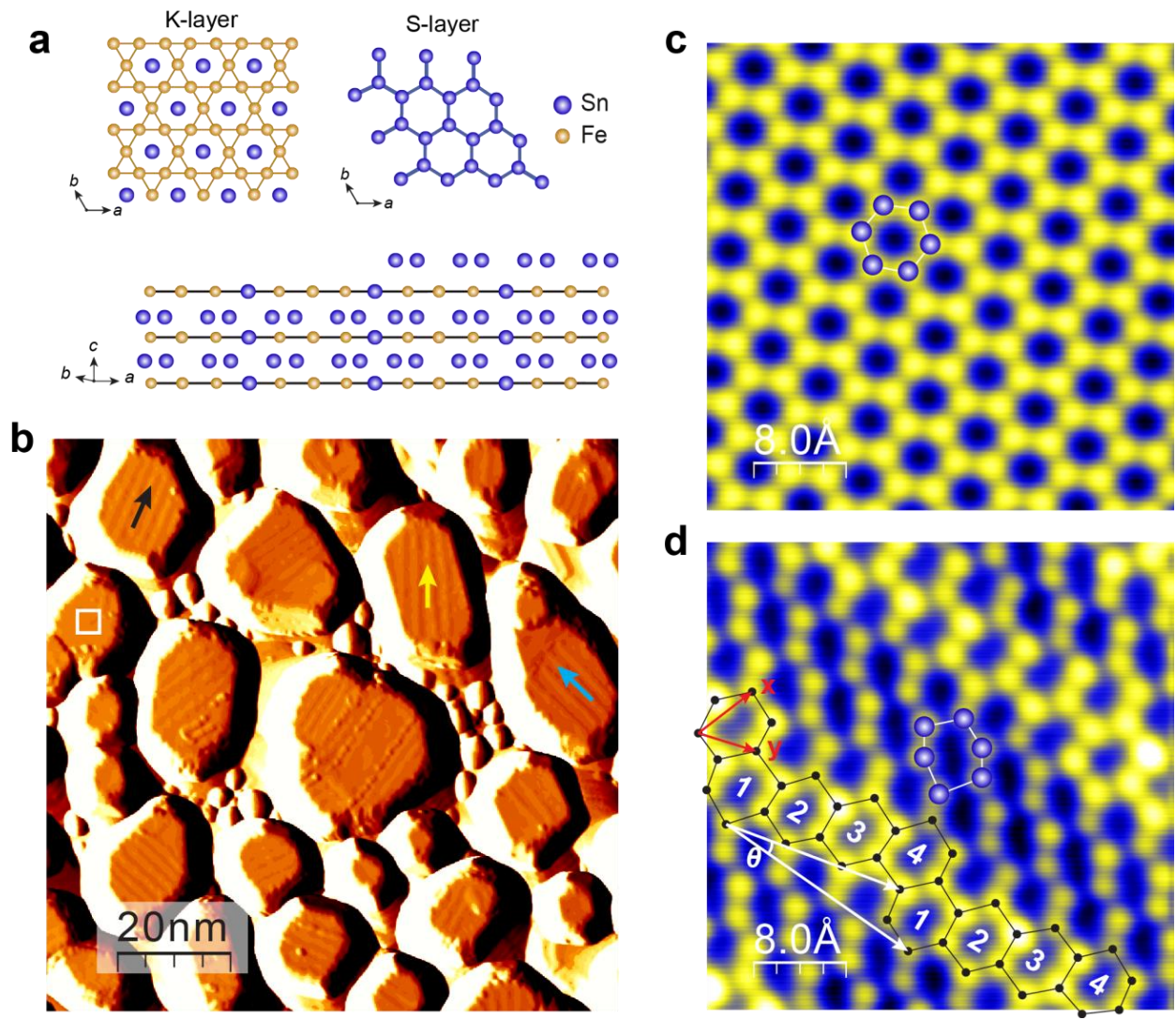

**Figure 1. Molecular beam epitaxy growth of FeSn films on SrTiO$_3$(111). a**, Ball-and-stick model of the FeSn crystal structure from top and side views. **b**, Morphology of a FeSn film grown at $T_{sub}$ = 480 °C, setpoint: $V$ = 3.0 V, $I$ = 10 pA. The STM image is in differential mode and the height of islands varies from to 6 to 7 nm. **c**, Atomic resolution STM image showing a perfect honeycomb lattice on the surface of a flat FeSn island, setpoint: $V$ = -2.0 mV, $I$ = 5.0 nA. **d**, Atomic resolution image revealing strongly distorted Sn honeycombs on the island with stripe modulations, setpoint: $V$ = 20 mV, $I$ = 5.0 nA.

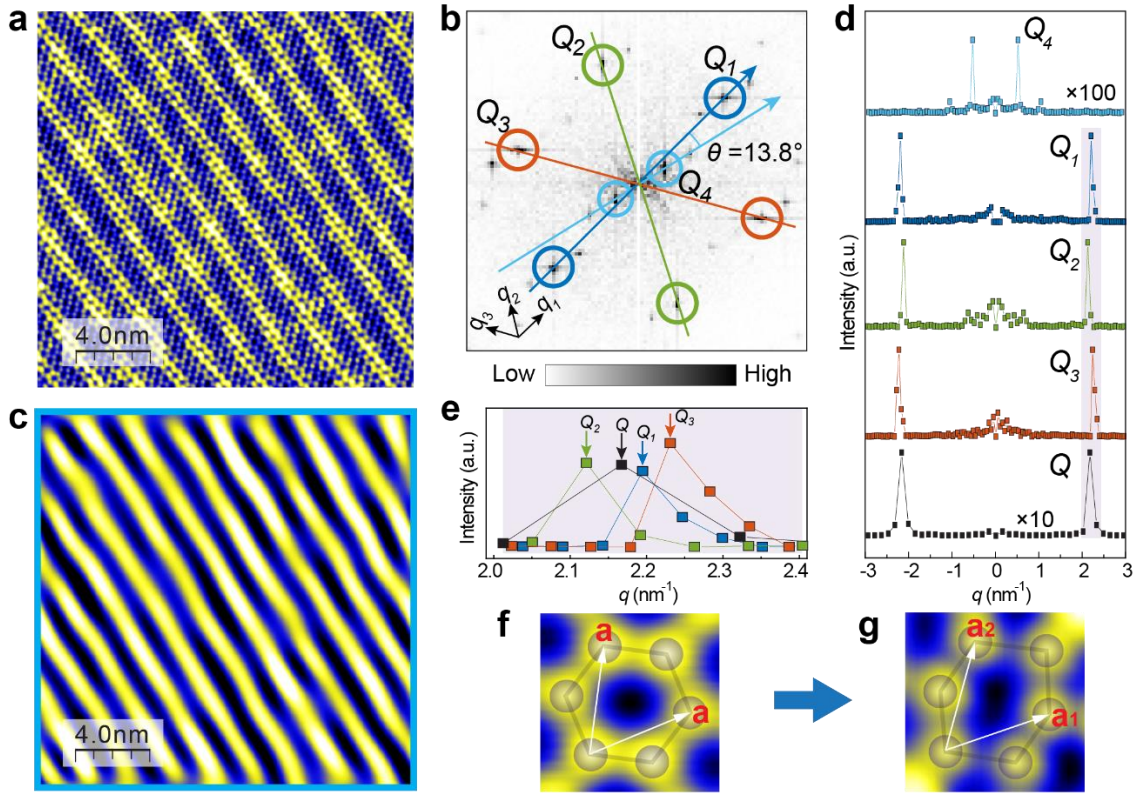

**Figure 2. FFT analysis of the stripe modulations in strained FeSn/SrTiO$_3$(111) films. a**, STM topographic image of stripe modulations of Sn honeycomb lattice, setpoint: $V$ = 100 mV, $I$ = 5.0 nA. **b**, FFT of the image in (**a**). Bragg peaks along three directions **$q_1$**, **$q_2$**, and **$q_3$** directions are denoted by $Q_1$, $Q_2$, and $Q_3$ in dark cyan, green and red circles, respectively. The diffraction peak of the stripe modulation is denoted by $Q_4$ in cyan circles. There is an angle of 13.8° between $Q_4$ and $Q_1$ directions. **c**, Reverse-FFT of the peak $Q_4$ in (**b**). The distribution of the stripes coincides with the topography in (**a**). **d**, Line profiles along the $Q_1$, $Q_2$, $Q_3$ and $Q_4$ directions in (**b**). As a reference, the line profile across the Bragg lattice $Q$ is obtained from the FFT of a perfect honeycomb lattice. **e**, Close-up view of the Bragg peaks $Q_1$, $Q_2$, $Q_3$, and $Q$. The peak position is marked by arrows in corresponding colors. **f**, Perfect honeycomb with lattice vector **a** marked. **g**, Deformed honeycomb with lattice vectors **$a_1$** and **$a_2$** marked.

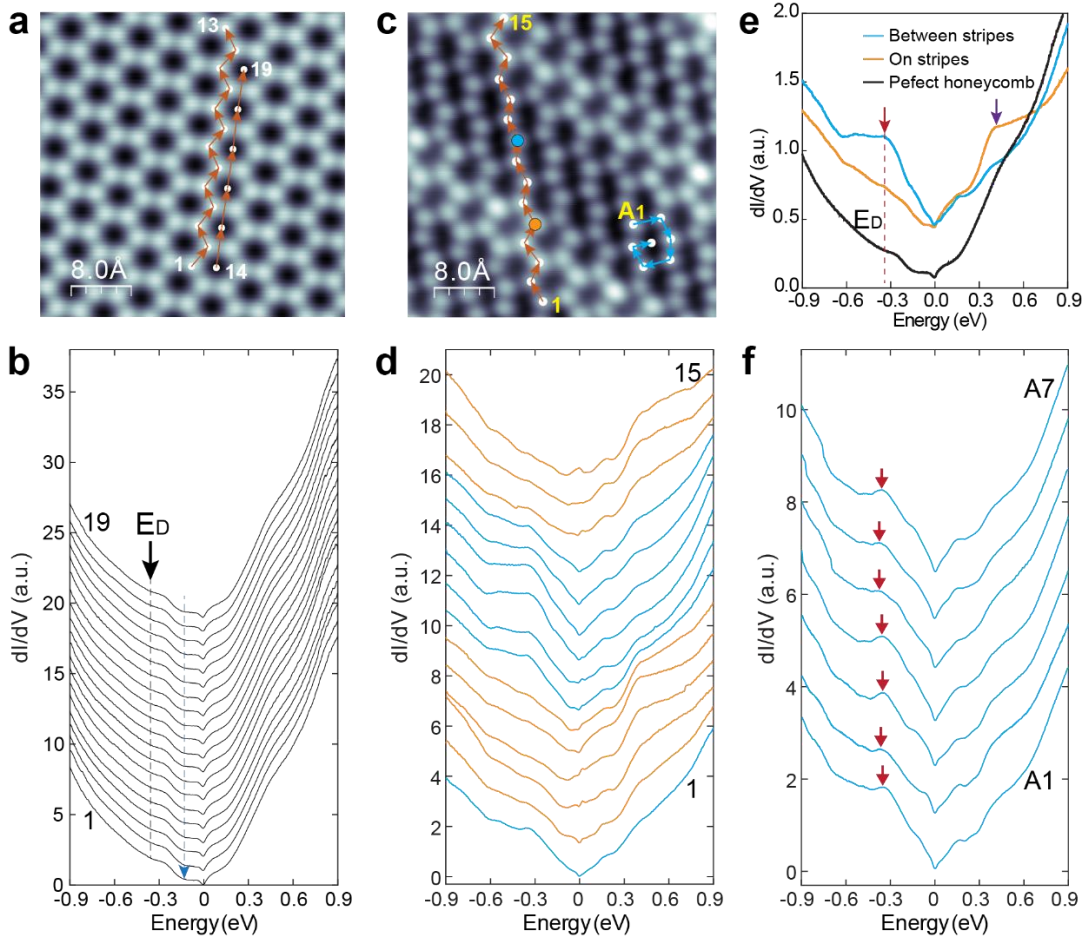

**Figure 3. Modulations of the electronic structure in strained FeSn/SrTiO$_3$(111) films. a**, Atomic resolution image of perfect honeycomb lattice, setpoint: $V$ = 3.0 V, $I$ = 10 pA. **b**, A series of dI/dV spectra taken at sites marked in (**a**). The energy position of the Dirac point $E_D$ is marked by the black arrow and dashed line, and the dip feature at -0.15 eV is marked by a cyan arrow. **c**, Atomic resolution image of distorted Sn honeycomb with stripe modulations, setpoint: $V$ = 10 mV, $I$ = 1.0 nA. **d**, A series of dI/dV spectra taken at sites marked in (**c**). The dI/dV spectra are classified into two groups, orange (taken on stripes) and cyan (taken between stripes). **e**, dI/dV spectra taken at sites on- (orange) and between stripes (cyan), compared to the reference spectrum taken on perfect Sn honeycomb without stripe modulations. The pronounced peak at $E$ = -0.36 eV (denoted with red arrow and dashed line) coincides with the Dirac point. **f**, dI/dV spectra taken within one-unit-cell of the strongly deformed honeycomb lattice, marked in (**c**). The energy

position of the $E_D$ is uniform within the deformed honeycomb lattice. The setpoint remains the same during the line dI/dV measurements.

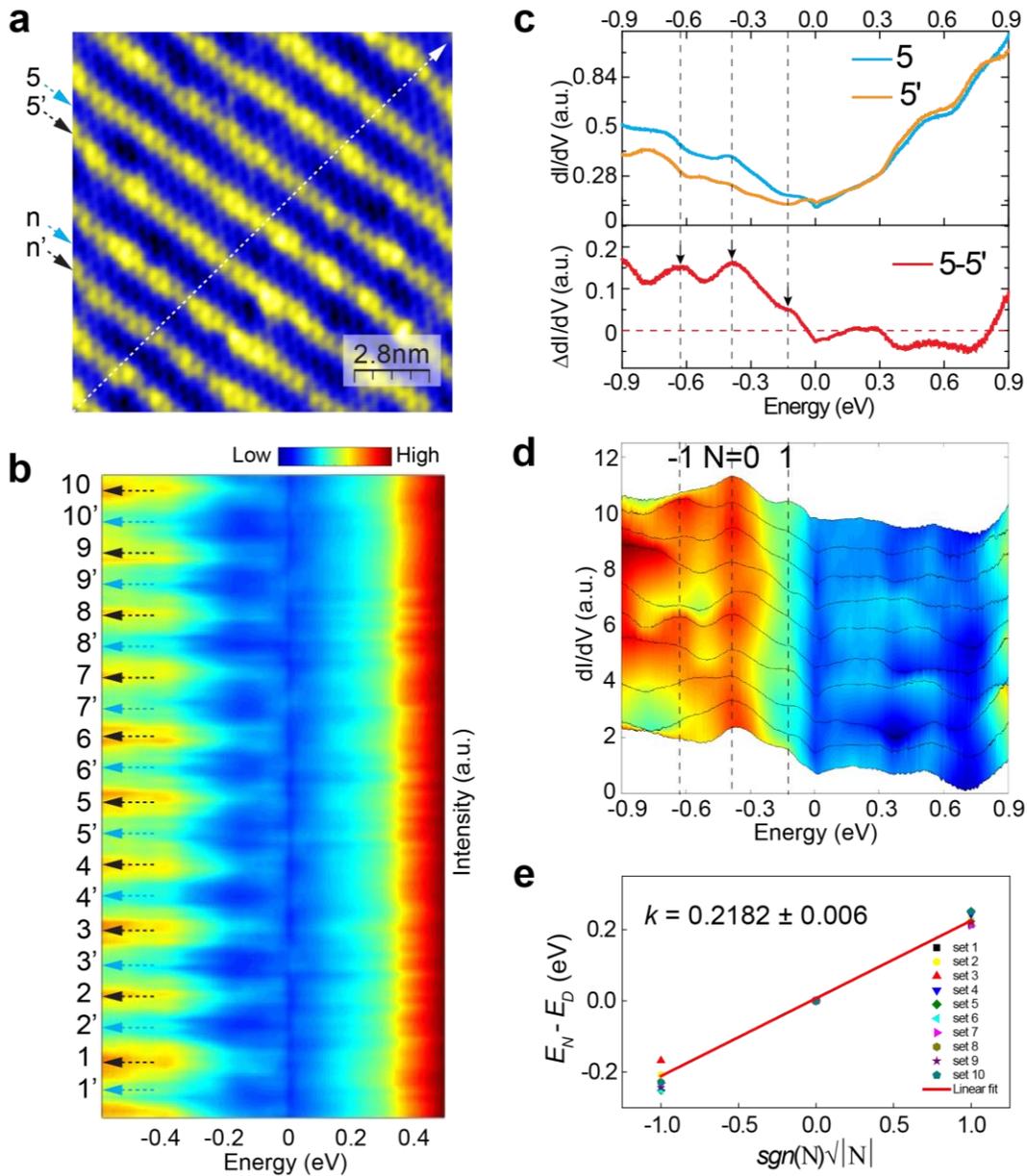

**Figure 4. Periodic pseudo-magnetic fields induced by stripe modulations in strained Sn-terminated FeSn/SrTiO$_3$(111) films. a**, STM topographic image showing the stripe modulations, where the on- and between stripe sites are labeled as n' and n, respectively. Setpoint: $V$ = 1.0 V, $I$ = 7.0 nA. **b**, dI/dV spectra taken along the white arrow in (**a**) with the intensity displayed in false color. The setpoint remains the same during the line dI/dV measurements. **c**, Upper panel: dI/dV

spectra taken on stripe (labeled 5' in (**a**)) and between stripes (5 in (**a**)). Lower panel: difference spectrum with three peaks marked by black arrows. **d**, Difference dI/dV spectra by the subtracting n and n'. The most pronounced peak at -0.36 eV is assigned to $N = 0^{th}$ Landau level and the neighboring two peaks are assigned to $N = \pm 1^{st}$. **e**, Plot of peak energy ($E_N - E_D$) vs. $sgn(N)\sqrt{|N|}$. The linear fitting (red line) yields to a slope $k = 0.2182 \pm 0.006$.

**Table of content**

**By scanning tunneling microscopy/spectroscopy**, we report giant periodic pseudo-magnetic field larger than 1000 Tesla induced by the periodic stripe modulations in epitaxial kagome magnet FeSn thin films on SrTiO$_3$(111) substrate.

*Huimin Zhang, Michael Weinert, and Lian Li\**

**Giant periodic pseudo-magnetic fields in strained kagome magnet FeSn thin films on SrTiO$_3$(111) substrates**

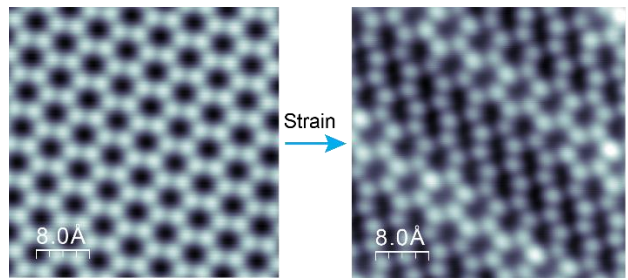

# Supporting Information

**Giant periodic pseudo-magnetic fields in strained kagome magnet FeSn epitaxial films on SrTiO$_3$(111) substrate**

Huimin Zhang[1,2], Michael Weinert[3], and Lian Li[1*]

[1]Department of Physics and Astronomy, West Virginia University, Morgantown, WV 26506, USA

[2]State Key Laboratory of Structural Analysis, Optimization and CAE Software for Industrial Equipment, Dalian University of Technology, Dalian, 116024, China

[3]Department of Physics, University of Wisconsin, Milwaukee, WI 53201, USA

[*]Correspondence to: lian.li@mail.wvu.edu

**This file includes:**

Note 1: MBE growth of FeSn films on SrTiO$_3$(111) substrate

Note 2: Calculations of the pseudo-magnetic field

Note 3: Modulation of electronic properties by stripes revealed by dI/dV maps

Figures S1-S21

**Note 1: MBE growth of FeSn films on SrTiO$_3$(111) substrate**

The FeSn films were grown by molecular beam epitaxy (MBE) on Nb-doped (0.05 wt.%) SrTiO$_3$(111) substrates. The SrTiO$_3$(111) substrates were first degassed at 600 °C for 3 hours and then followed by annealing at 950 °C for 1 hour to obtain a flat surface with step-terrace morphology. During the MBE growth, high purity Fe (99.995%) and Sn (99.9999%) were evaporated from Knudson cells setting at 1150 °C and 805 °C and co-deposited on the SrTiO$_3$(111) substrate at a substrate temperature ($T_{sub}$) between 480 to 530 °C. The Fe/Sn ratio is estimated to be 1:2.7 during growth. The substrate temperature is monitored by a LAND portable infrared thermometer (Cyclops 160L). At $T_{sub}$ = 480 °C, the surface of the film exhibits stripe modulations while at elevated $T_{sub}$ = 530°C, such modulations are mostly absent, as shown in Figure S3. Multiple samples (> 5) were grown with these parameters.

FeSn consists of alternating Sn$_2$ honeycomb (Sn termination) and Fe$_3$Sn kagome layer (Fe$_3$Sn termination) vertically stacked. The stripe modulation only appears on Sn termination, while the Fe$_3$Sn termination exhibits the close-packed structure without noticeable distortion (Figure S5). The striped surface shows crest and trough morphology with a height contrast ~20.9 pm (c.f., Figure S10 for bias-dependent behavior). At atomic scale, the valley region (on the stripe) is of slight distortion while the trough region exhibits strong distortion, as shown with the ball-and-stick model imposed on top (Figure 1**g**).

**Note 2: Calculations of the pseudo-magnetic field**

In general, the response of Dirac materials, i.e., materials whose low energy excitations are described by massless Dirac equation, to strain is well understood[1]. For a honeycomb lattice, strain induced distortion will lead to axial vector and scalar potentials, which will strongly modify its electronic properties[2]. At the level of tight binding model, the potential will modify the Dirac Hamiltonian:

$$-i\hbar v_F \vec{\sigma}(\nabla - i\vec{A})$$

where $A_x = \frac{\sqrt{3}}{2}(t_3 - t_2)$ and $A_y = \frac{1}{2}(t_2 + t_3 - 2t_1)$, and $t_i$ are the nearest neighbor hoping parameters. For a perfect honeycomb lattice, the $t_i$ are the same, resulting in zero A. For distorted honeycomb, however, the lattice distortion will modify the lattice constant, thus the hopping parameter, leading to non-zero A. Furthermore, assuming that the atomic displacements $u_i$ are small in comparison with the lattice constant, then the hopping parameter can be written as:

$t_i = t + \frac{\beta t}{a^2} \vec{\delta}_i (\vec{u}_i - \vec{u}_0)$ and $(\vec{u}_i - \vec{u}_0) \propto (\vec{\delta}_i \nabla) \vec{u}(\vec{r})$ ($\vec{\delta}_i$ is the nearest neighbor vectors).

This leads to an effective magnetic field $B_{eff} = \nabla \times \vec{A}$, with

$$A_x = c\frac{\beta t}{a}(u_{xx} - u_{yy}), A_y = -c\frac{2\beta t}{a}u_{xy}$$

This effective field will result in Landau levels (LLs) that are expected to follow the equation, same as the one caused by the application of real electromagnetic fields[3,4]:

$$E_N = E_D + sgn(N)v_F\sqrt{2e\hbar|N|B_{eff}} \qquad (1)$$

Here, $E_D$ is the energy of Dirac point, $v_F$ is the Fermi velocity, $e$ is the electron charge $1.602 \times 10^{-19}$ C, $\hbar$ is the reduced Planck constant $1.0546 \times 10^{-34}$ J·s, and $B_{eff}$ is the pseudo-magnetic field. The integer $N$ represents an electron-like ($N$ > 0) or a hole-like ($N$ < 0) LL index.

$$k = \frac{E_N - E_D}{sgn(N) \cdot \sqrt{N}} \qquad (2)$$

The slope $k$ is obtained from experimental data, as shown in Fig. 4e where $E_N - E_D$ is plotted as a function of $sgn(N)\sqrt{|N|}$. The linear fit in Fig. 4e yields to a slope $k$ = 0.2182 ± 0.006 eV.

$$B_{eff} = \frac{k^2}{2e\hbar v_F^2} \qquad (3)$$

Basing on Equation 3, we can calculate the effective pseudo-magnetic field $B_{eff}$. If we take the bulk FeSn value $v_F = (1.7 \pm 0.2) \times 10^5$ $ms^{-1}$ due to lack of data for epitaxial FeSn films, the pseudo-magnetic field is determined to be $B_{eff}$ = 1251 ± 363 T. If we take the Fermi velocity of stanene ($v_F = 4.4 \times 10^5$ $ms^{-1}$)[5], the effective pseudo-magnetic field $B_{eff}$ = 187 ± 10 T is obtained. In addition, the dI/dV spectra in Sn layer of FeSn (Fig. 3**b**) are intrinsically different from that of stanene on Au(111)[6] or Sb(111)[7]. Therefore, to adopt the Fermi velocity of FeSn is more appropriate.

The LLs for strain-induced pseudomagnetic field in kagome layer are expected to follow similar equation, however with an anisotropic Fermi velocity[8].

**Note 3: Modulation of electronic properties by stripes revealed by dI/dV maps**

The Sn honeycomb with stripe modulations are investigated by dI/dV mapping. The topographic image and its corresponding spatial dI/dV maps are displayed in Fig. S17**a**. Accompanied the stripes in the topography, we observe stripy electronic modulation across the surface in the dI/dV maps within the energy range [-500 meV, 500 meV]. Figure S17**b** compares the line profile derived from the topography and corresponding dI/dV maps, where the dashed red lines mark the valley position from the topography. Between the energy ranges [-500 meV, -100 meV] and [100 meV, 300 meV], the valley position in the topography corresponds to higher density of states. At Fermi level $g(\mathbf{r}, 0\text{ meV})$, the modulation is almost gone. At the energy window [400 meV, 500 meV], the valley position in the topography corresponds to lower density of states. In other words, there is a $\pi$-phase shift between 400 meV and 500 meV, compared to the other energies.

We notice that there are different stripe orientations for different FeSn islands (Figure S6). dI/dV spectra and dI/dV maps are carried out for stripes with different orientations and we find that the orientation of the stripes has no detectable effect on the dI/dV spectra or maps. As shown in Figure S19, we find that not only the dI/dV spectra but also the spatial modulation is consistent along different orientations.

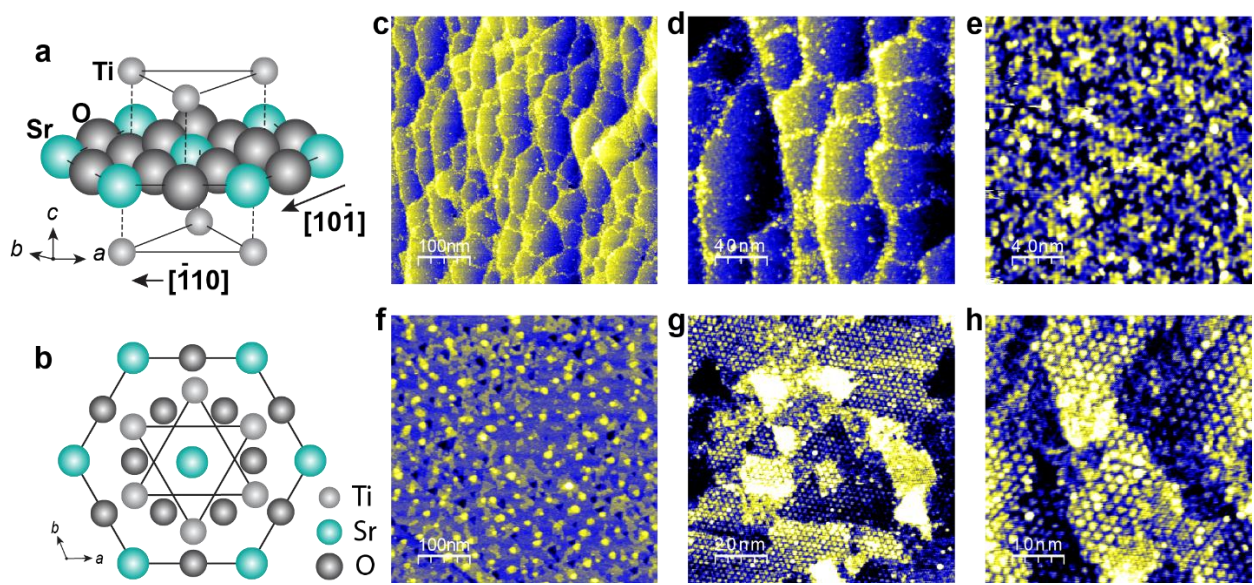

**Figure S1. STM imaging of the annealed SrTiO$_3$(111) substrate.** Ball-and-stick model of the SrTiO$_3$(111) surface: side view (**a**) and top-view (**b**). **c-e**, Topographic STM images of type-1 SrTiO$_3$(111) substrate. Setpoint: $V$ = 3.0 V, $I$ = 30 pA (c)-(d), and $V$ = 1.0 V, $I$ = 500 pA (**e**). **f-h**, Topographic STM images of type-2 SrTiO$_3$(111) substrate. Setpoint: $V$ = 2.0 V, $I$ = 10 pA for (**f**)-(**g**), and $V$ = 2.0 V, $I$ = 30 pA for (**h**). The SrTiO$_3$(111) surface exhibits a (4 × 4) reconstruction. The growth of FeSn films is similar on both types of SrTiO$_3$(111) substrates.

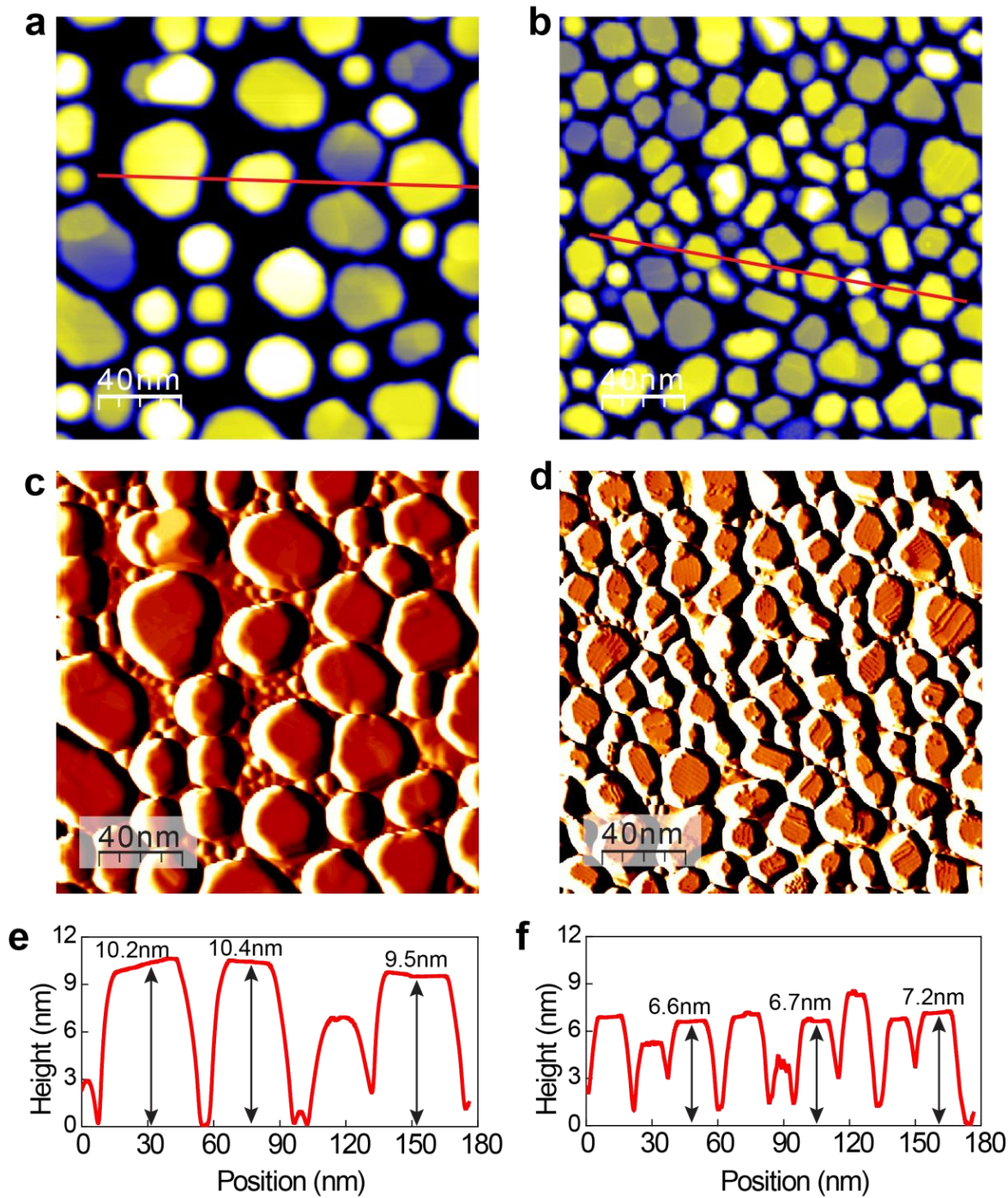

**Figure S2. Initial stages of FeSn growth on SrTiO$_3$(111) substrates. a-b**, STM topographic images of FeSn/SrTiO$_3$(111) films. Setpoint: *V* = 2,0 V, *I* = 10 pA. **c-d**, STM images of (**a**) and (**b**) in differential mode. **e-f**, Line profiles along the red line in (**a**) and (**b**), respectively.

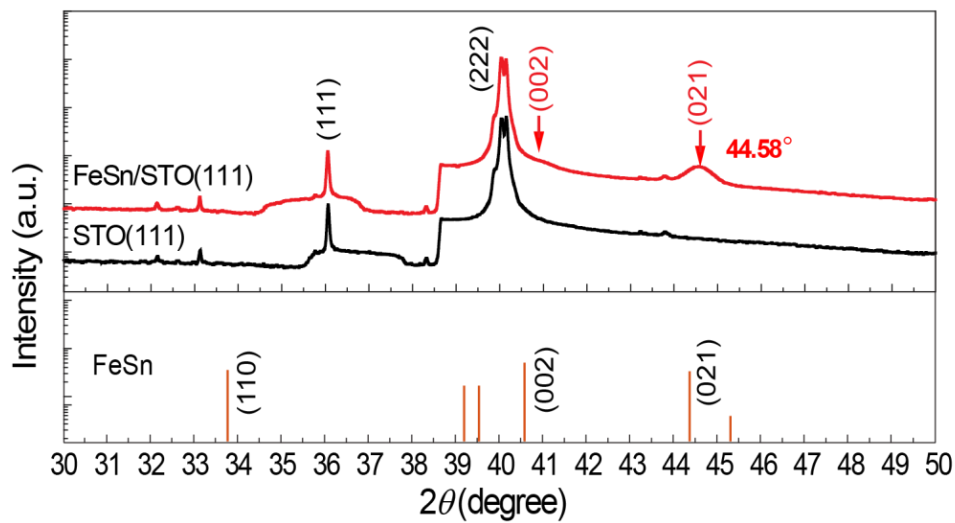

**Figure S3. X-ray diffraction (XRD) data of a 20 nm FeSn/SrTiO$_3$(111) film.** The peaks at $2\theta$ = 40.52° and 44.58° are indexed as FeSn (002) and (021), confirming the FeSn phase. Note that both peaks are slightly shifted to higher values, indicating smaller lattice constant in the *c*-direction.

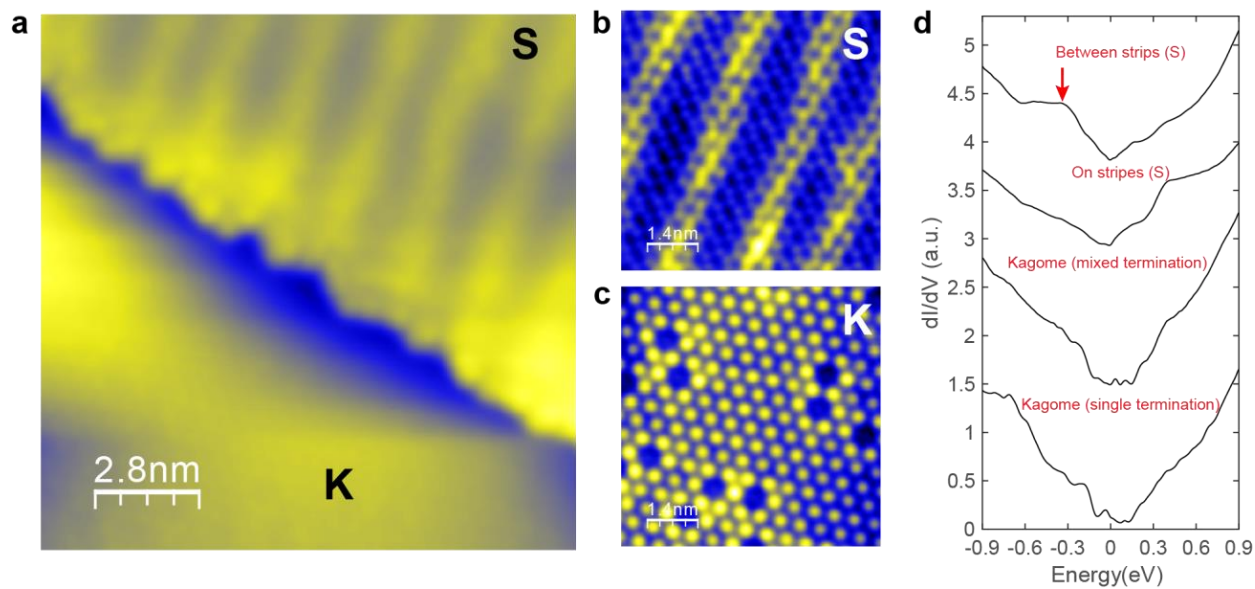

**Figure S4. STM imaging of the mixed Sn termination (S layer) and Fe$_3$Sn kagome termination (K layer) in FeSn/STO(111) films. a**, Topographic STM image showing both S and K terminations. Setpoint: $V$ = 0.9 V, $I$ = 3.0 nA. **b**, The STM image of the S-layer with stripe modulations. Setpoint: $V$ = -0.1 V, $I$ = 3.0 nA. **c**, STM image of the K-layer without stripe modulations. Setpoint: $V$ = 0.2 V, $I$ = 1.0 nA. **d**, Typical dI/dV spectra taken between stripes and on stripes on the Sn layer; and on single and mixed K layers.

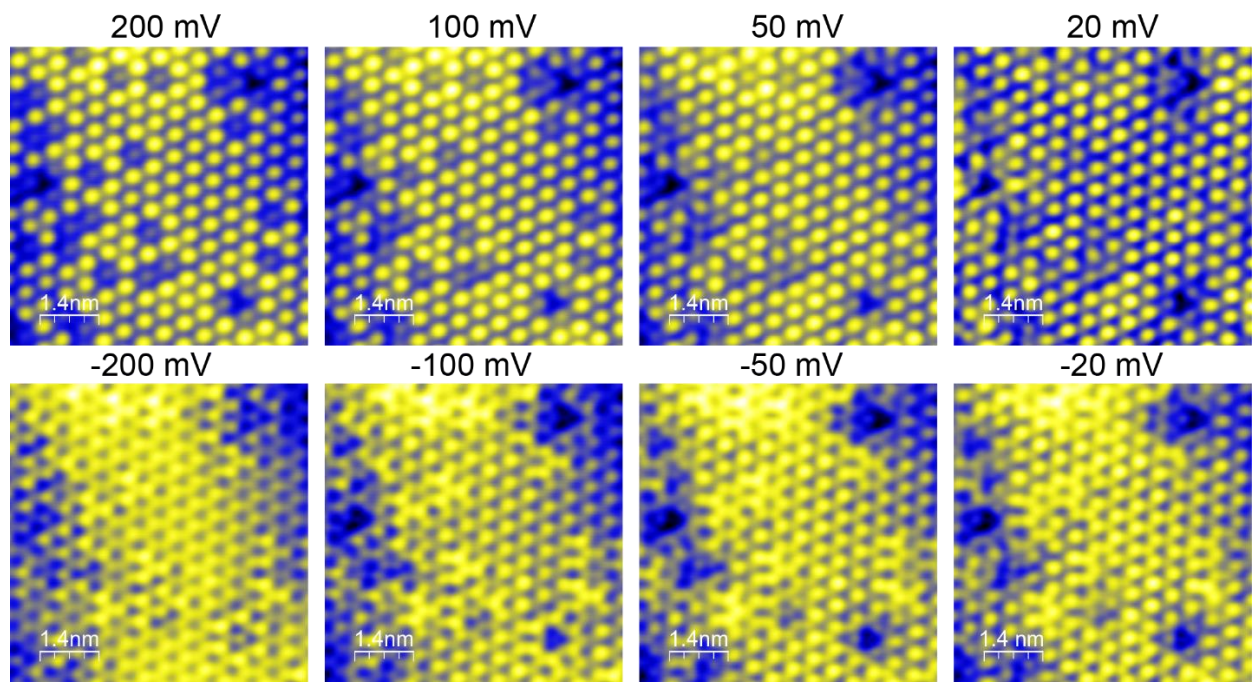

**Figure S5. Bias-dependent STM imaging of the kagome layer on the K-terminated FeSn/SrTiO$_3$(111) film.** Setpoint: $I$ = 5.0 nA with $V$ specified.

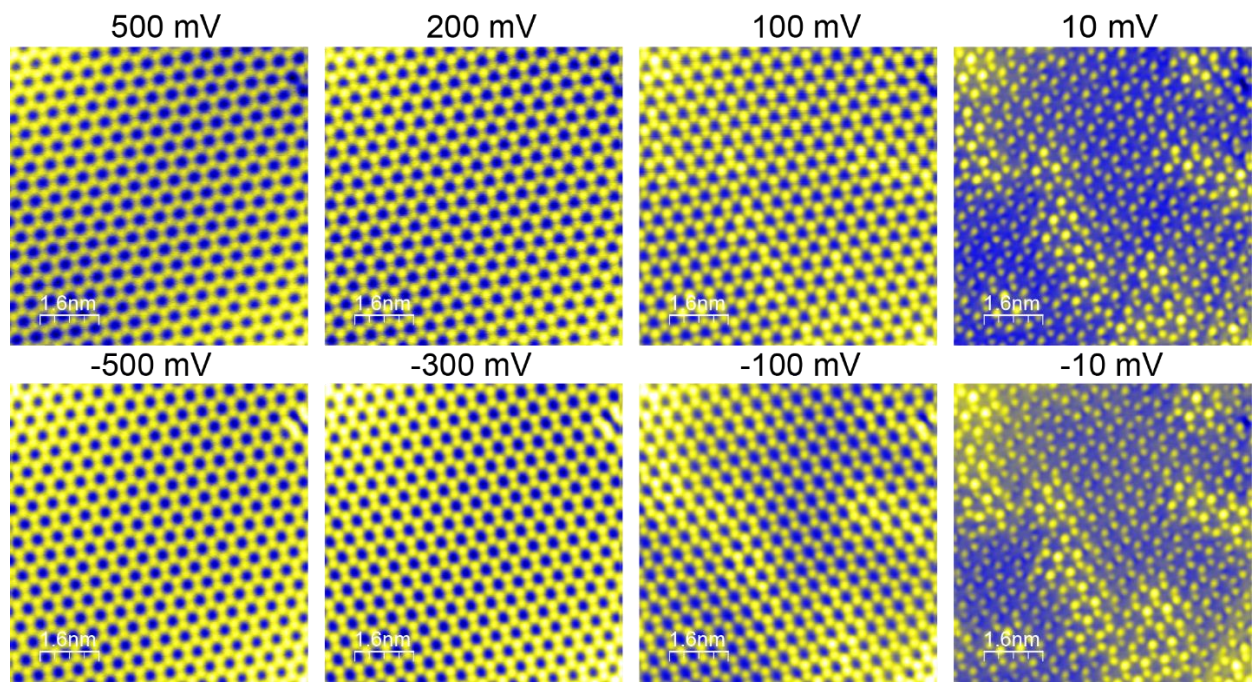

**Figure S6. Bias-dependent STM imaging of the Sn honeycomb on the Sn terminated FeSn/SrTiO$_3$(111) film.** Setpoint: *I* = 5.0 nA with *V* specified. The defect states are revealed at -10 and 10 mV.

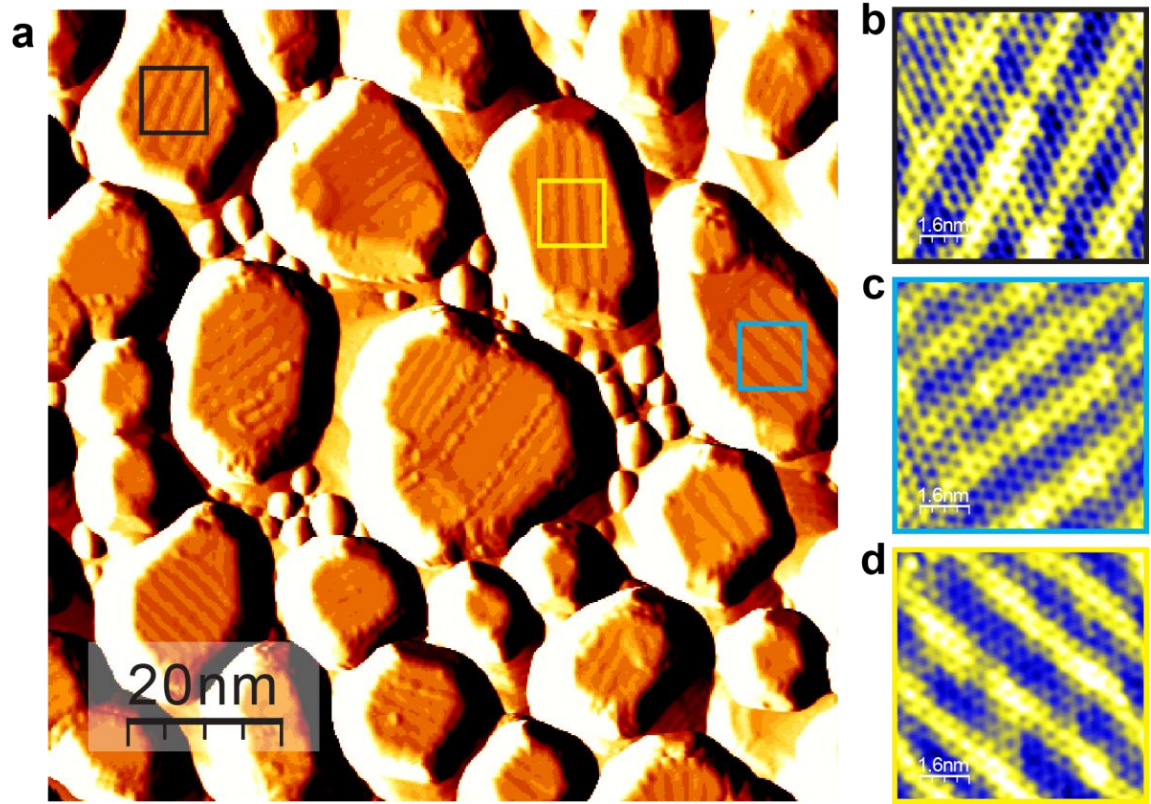

**Figure S7. Stripe modulations on Sn terminated FeSn/SrTiO₃(111) films. a**, STM image of a FeSn film grown on SrTiO₃(111) substrate in differential mode, setpoint: $V$ = 1.0 V, $I$ = 100 pA. **b-d**, STM images of the stripe modulations taken in the regions outlined by black, cyan, and yellow squares in (**a**). Setpoint: $V$ = 3.0 mV, $I$ = 5.0 nA.

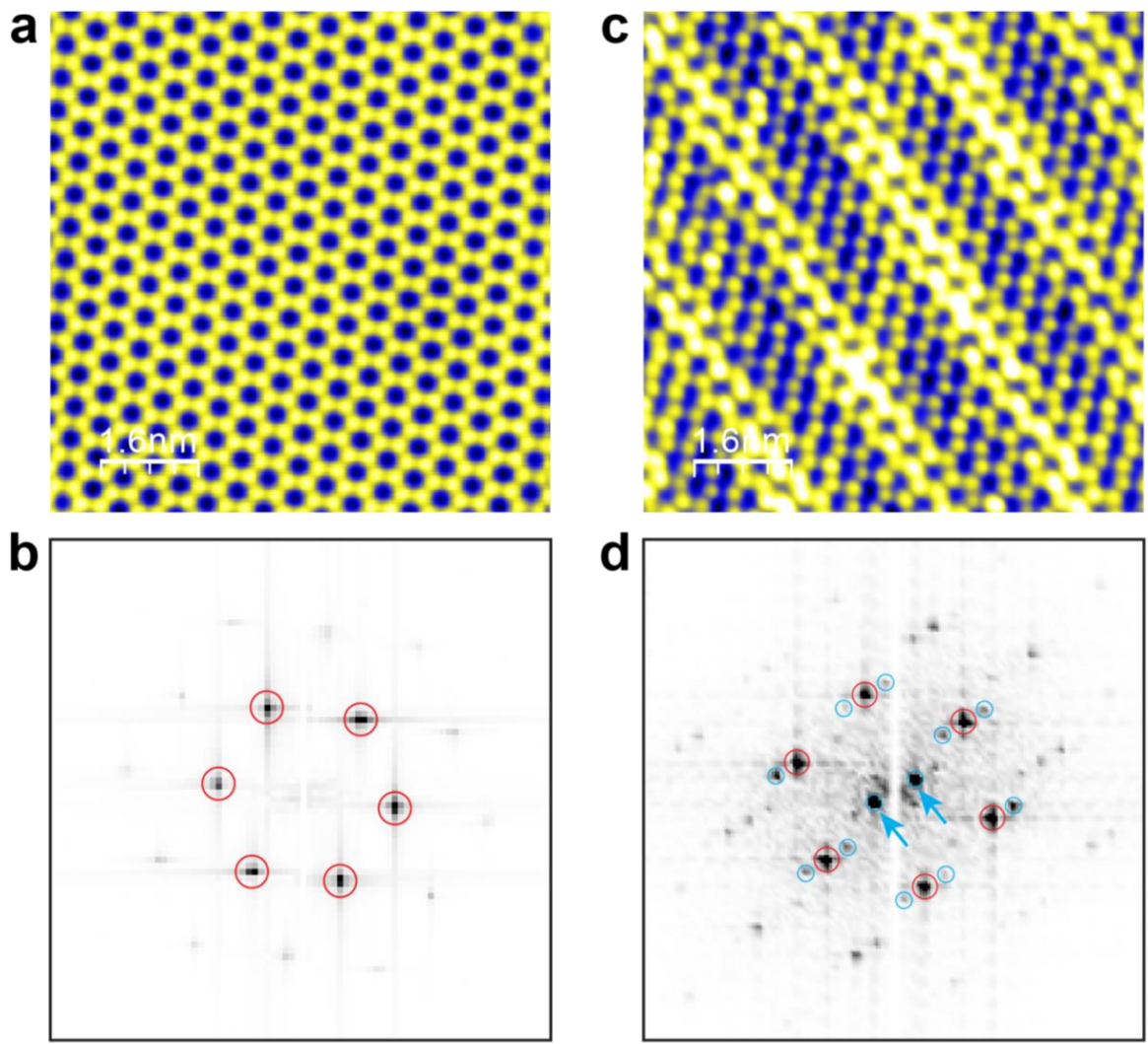

**Figure S8. FFT of perfect and deformed honeycomb lattices. a**, Atomically resolved STM image of the deformed honeycomb lattice, setpoint: $V$ = 100 mV, $I$ = 5.0 nA. **b**, FFT of (**a**) with Bragg peaks $Q_1$, $Q_2$ and $Q_3$ marked in red circles. The $Q_4$ in cyan circle denotes the stripe modulation. **c**, Atomic resolution image of a perfect honeycomb lattice, setpoint: $V$ = -2.0 mV, $I$ = 5.0 nA. **d**, FFT of (**c**). The red circles mark the Bragg peak $Q$.

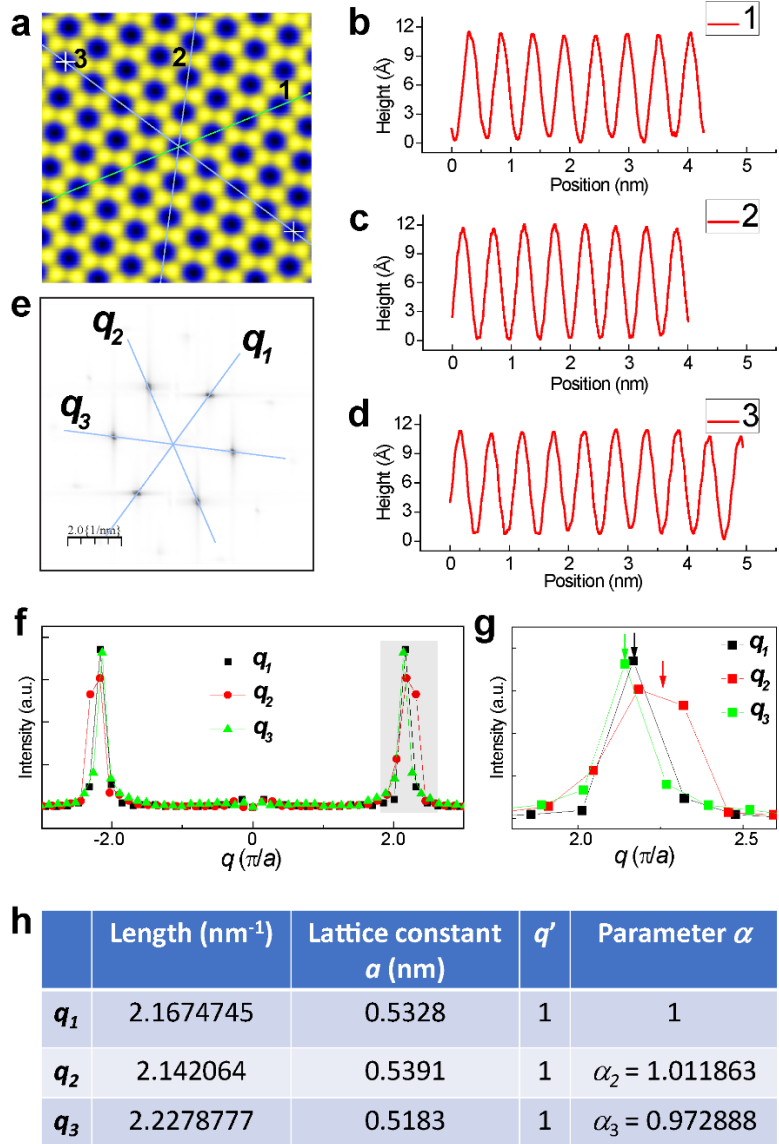

| | Length (nm$^{-1}$) | Lattice constant $a$ (nm) | $q'$ | Parameter $\alpha$ |
|---|---|---|---|---|
| $q_1$ | 2.1674745 | 0.5328 | 1 | 1 |
| $q_2$ | 2.142064 | 0.5391 | 1 | $\alpha_2$ = 1.011863 |
| $q_3$ | 2.2278777 | 0.5183 | 1 | $\alpha_3$ = 0.972888 |

**Figure S9. Calibration of lattice distortion caused by nonlinearity of the STM scanner. a**, Atomic resolution image of honeycomb lattice, setpoint: $V$ = -2.0 mV, $I$ = 5.0 nA. **b**-**d**, Line profiles across the lines marked as 1, 2 and 3 in (**a**). **e**, FFT of (**a**). **f**, Line profiles along $q_1$, $q_2$ and $q_3$ directions marked in (**e**). **g**, Zoom-in of the shaded region in (**f**). **h**, Calibration of lattice constant by the formula $q' = \alpha * q$, where $q'$ represents the calibrated vector, and $q$ represents the measured lattice constants along three directions from STM imaging.

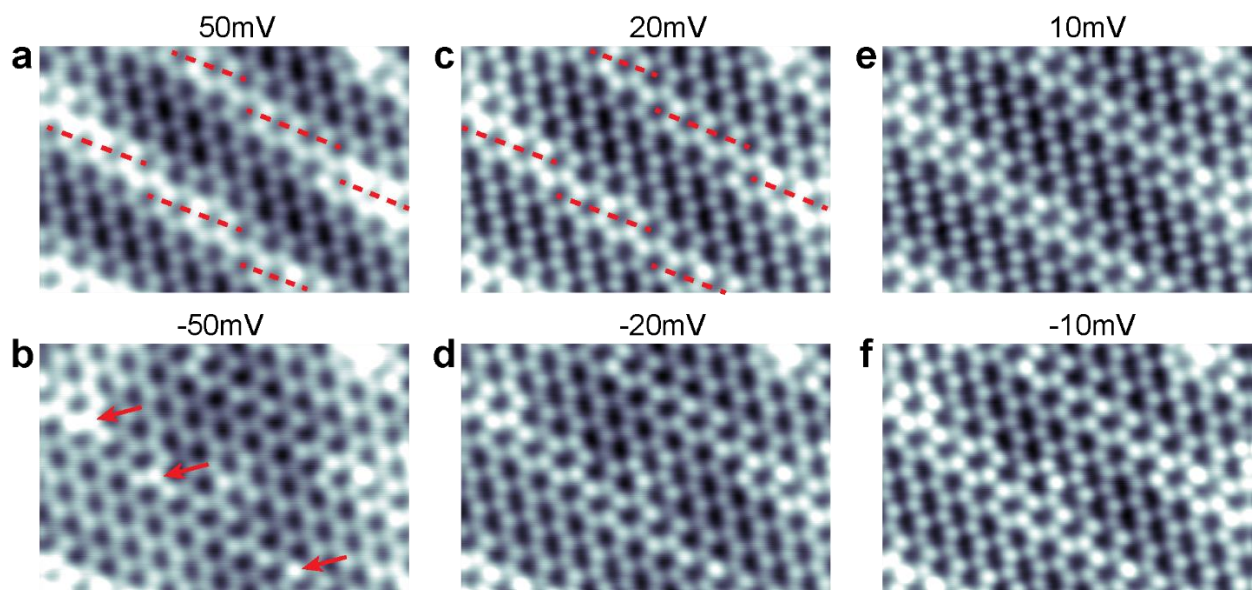

**Figure S10. Bias-dependent STM imaging of distorted honeycomb lattice.** The red dashed lines in (**a**) and (**b**) indicates the interlaced structure of the stripes. Setpoint: $I$ = 1.0 nA with bias $V$ specified. Size of the image: 4.0 nm × 6.0 nm. The red arrows in (**b**) mark the kink points of the stripes that display a higher contrast under $V$ = -50 mV.

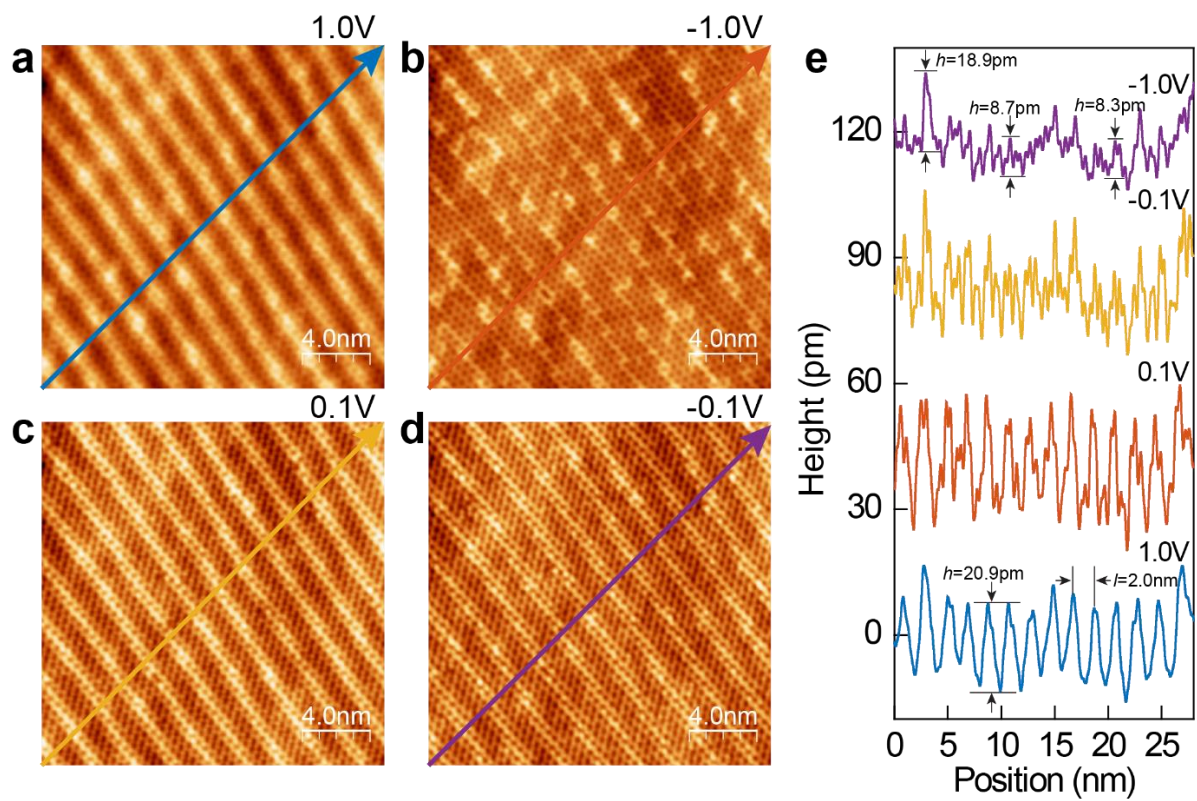

**Figure S11. Bias-dependence of stripe modulations. a-d**, Topographic STM images of stripe modulations under different bias *V*. Setpoint: *I* = 5.0 nA with *V* specified. **e**, Line profile across the topographic image under different bias *V* with the direction marked by arrows in (**a-d**). The high and low contrast under *V* = 1.0 V contributes to *h* = 20.9 pm and the periodicity cross the stripes is determined to be *l* = 2.0 nm.

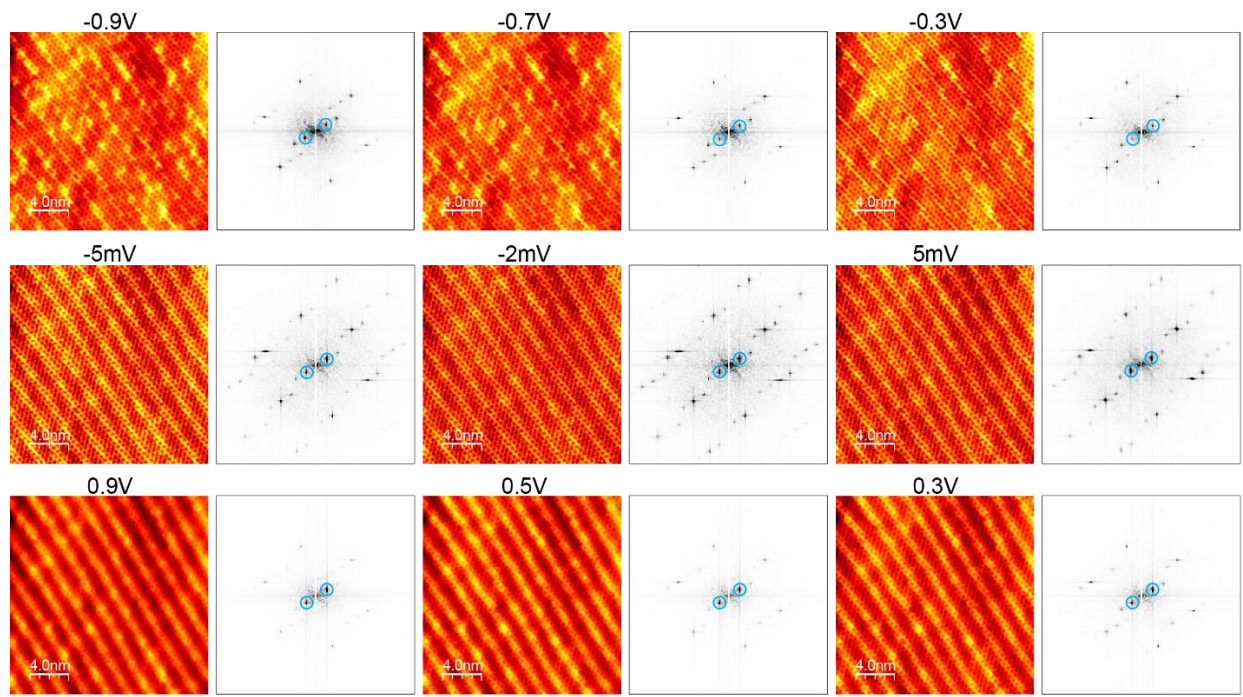

**Figure S12. FFT of bias-dependent STM images.** The STM images are taken with a tunneling current $I$ = 5.0 nA with $V$ specified. The diffraction peaks of the stripes are denoted by a pair of cyan circles in FFTs.

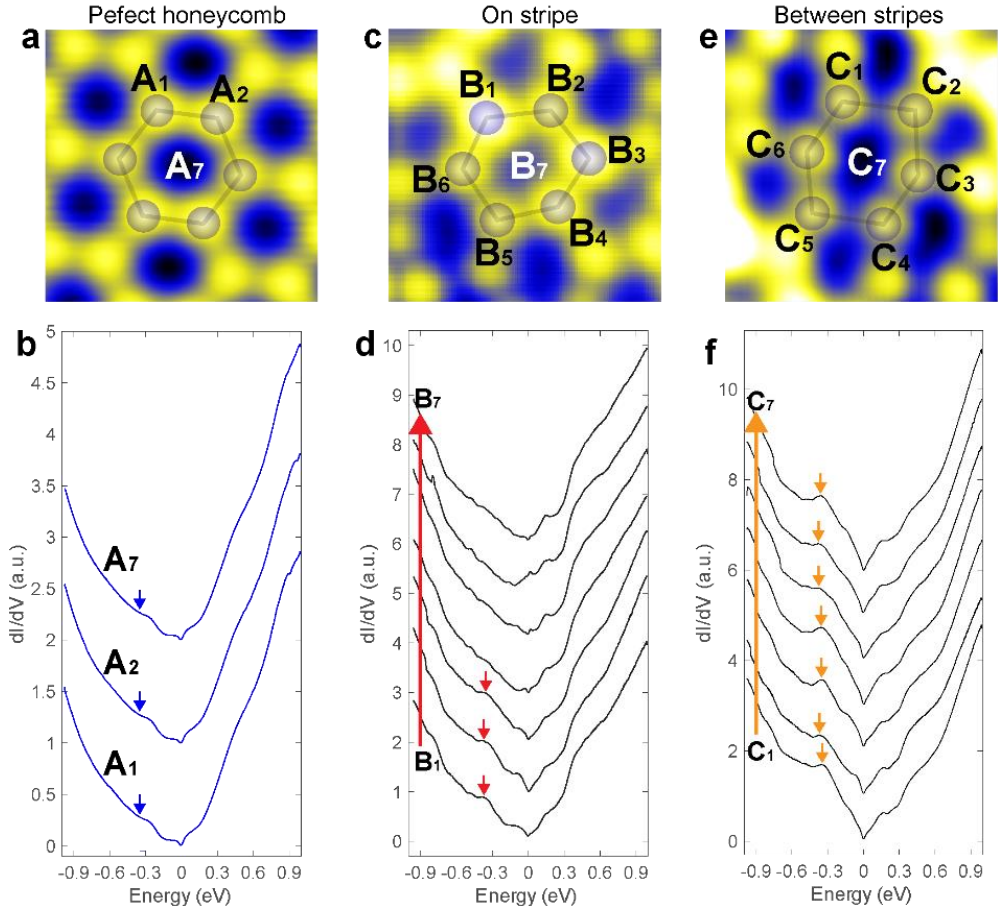

**Figure S13. Site-dependent dI/dV spectra within one honeycomb lattice. a**, Atomic resolution image showing perfect honeycomb lattice taken on the surface without stripe modulation, setpoint: $V$ = -2.0 mV, $I$ = 5.0 nA. **b**, dI/dV spectra at three typical sites, as labeled $A_1$, $A_2$ and $A_7$ in (**a**). The blue arrows denote the dip feature associated with the Dirac point $E_D$. **c**, STM image of the slightly deformed honeycomb on the stripe, setpoint: $V$ = 10 mV, $I$ = 1.0 nA. **d**, dI/dV spectra taken at sites labeled in (**c**). **e**, STM image of the strongly deformed honeycomb between the stripes, setpoint: $V$ = 10 mV, $I$ = 1.0 nA. **f**, dI/dV spectra taken at different sites labeled in (**e**). Typical sites of slightly and strongly deformed honeycomb lattice are labeled as $B_1$ to $B_7$ and $C_1$ to $C_7$, respectively.

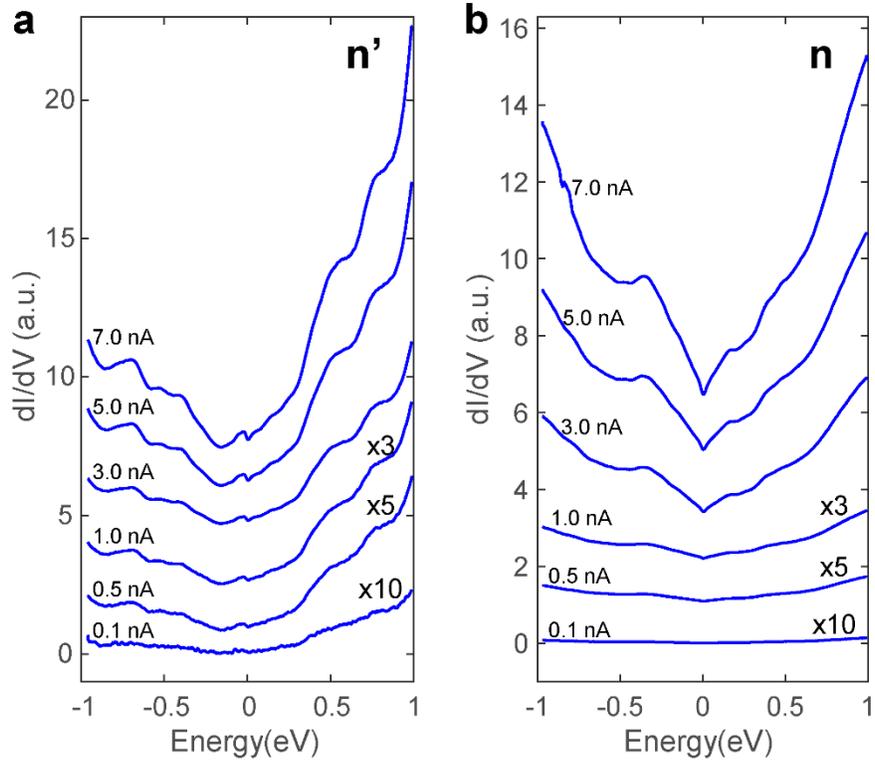

**Figure S14. Setpoint-dependent of dI/dV spectra on stripe (n') and between stripes (n).** Setpoint: $V$ = 1.0 V, with tunneling current $I$ specified. Spectra are vertically offset for clarity. The intensity of the dI/dV spectra taken at $I$ = 1.0, 0.5 and 0.1 nA are multiplied 3×, 5× and 10×, respectively.

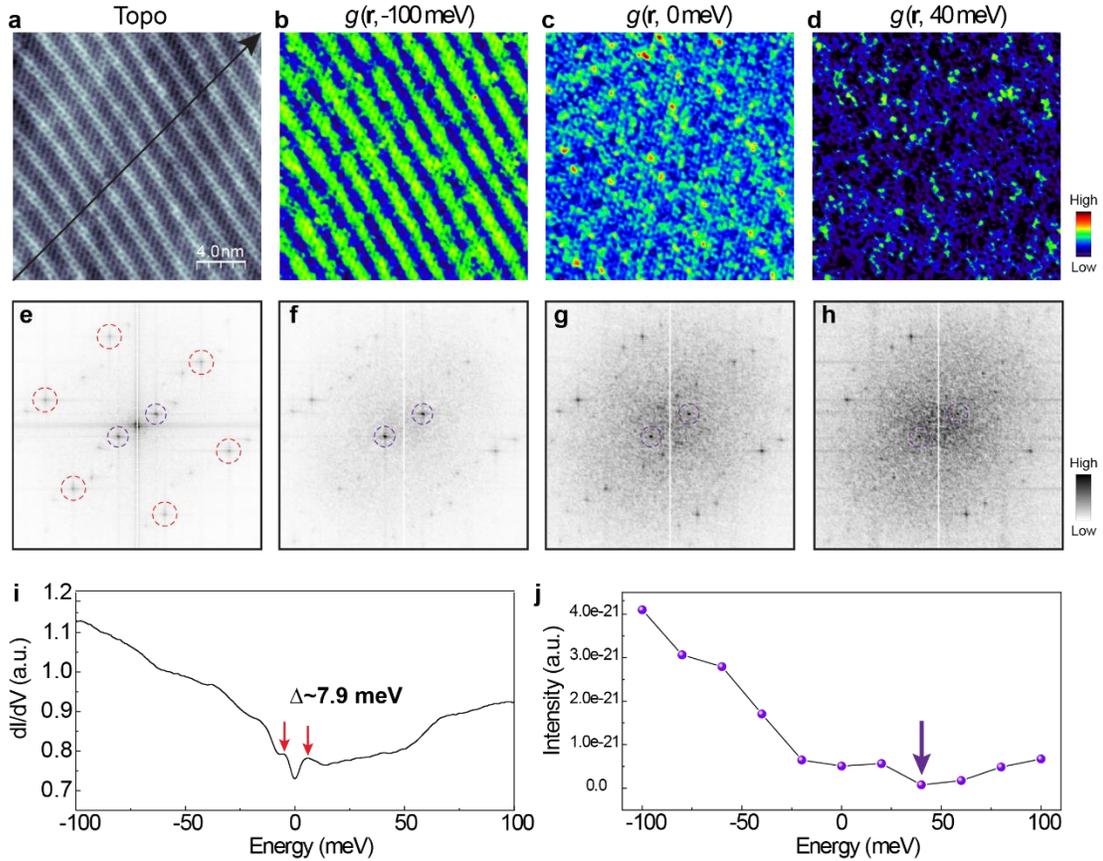

**Figure S15. FFT analysis of the stripe modulations. a**, Topographic STM image showing periodic stripe modulations. Setpoint: $V$ = 100 mV, $I$ = 5.0 nA. **b-d**, dI/dV maps of the region **a**. Setpoint: $V$ = 100 mV, $I$ = 5.0 nA, $V_{mod}$ = 2 meV. **e-h**, FFTs of **a-d**, respectively. The dashed red circles in **e** mark the Bragg lattice. The pair of dashed purple circles in **e-h** mark the stripe feature. **i**, dI/dV maps in the energy range [-100 meV, 100 meV]. The red arrows denote a gap feature $\Delta \sim 7.9$ meV at Fermi level. **j**, The intensity of the diffraction peak for the stripe modulation in FFTs as function of energy, where the minimal is located at 40 meV above the $E_F$.

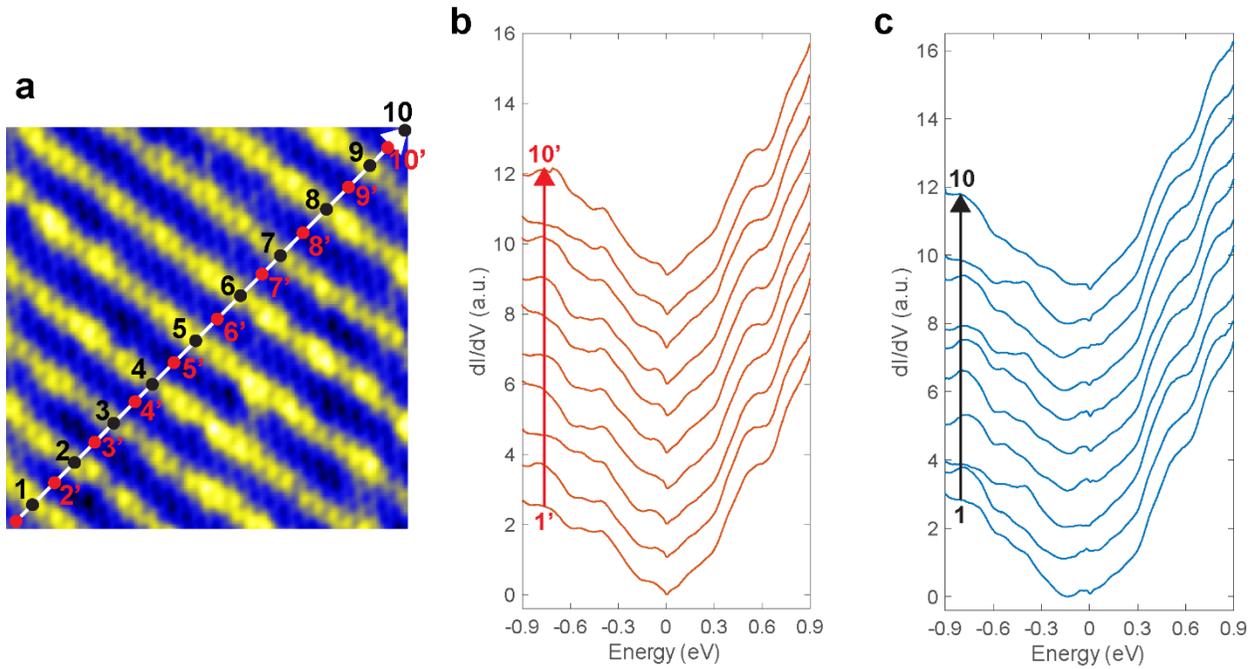

**Figure S16. Periodic stripe modulations. a**, Atomic resolution image of stripe modulations. Setpoint: $V$ = 1.0 V, $I$ = 7.0 nA. 1-10 label sites at the bright pattern (honeycomb lattice) and 1' to 10' label the sites between stripe patterns (deformed honeycomb lattice). **b**, dI/dV spectra at deformed honeycomb lattice, as labeled 1' to 10' in (**a**). **c**, dI/dV spectra at honeycomb lattice, as labeled 1 to 10 in (**a**).

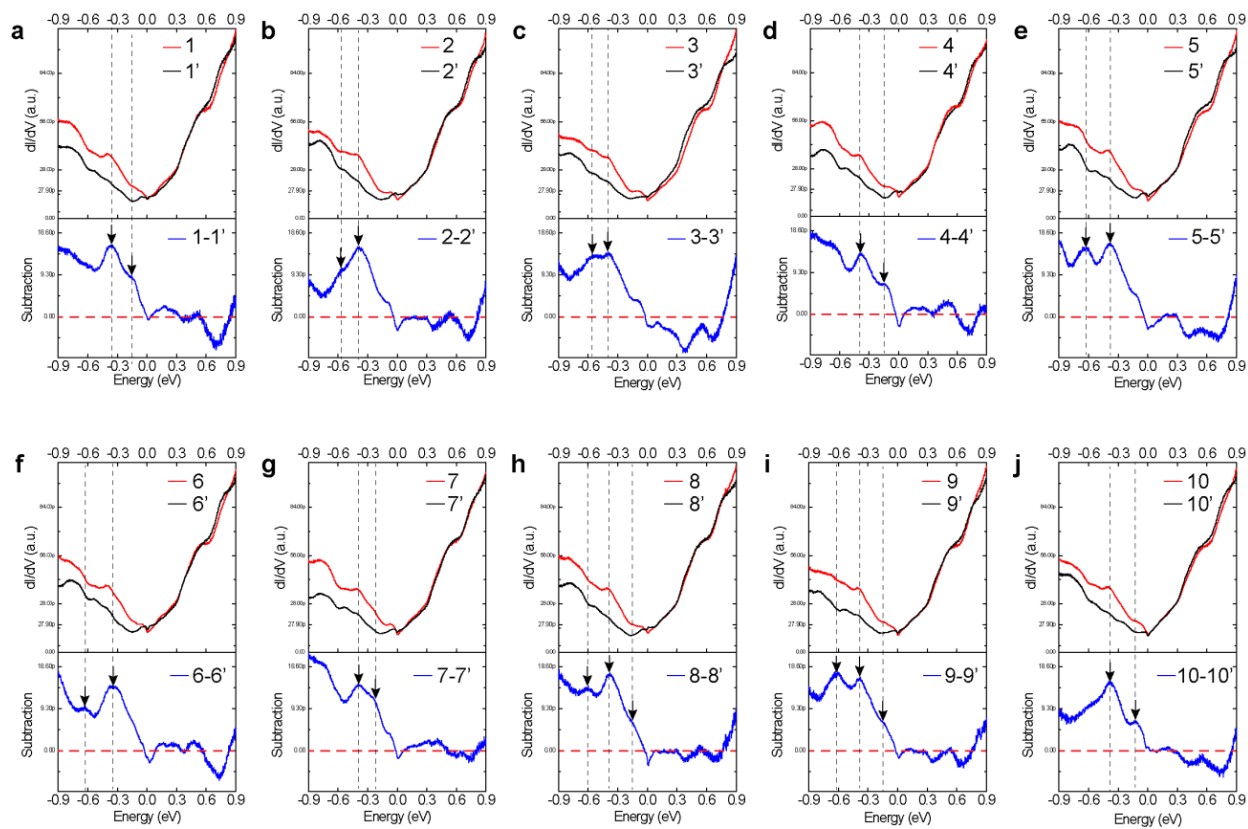

**Figure S17. Subtraction of 10 sets of dI/dV spectra to reveal the Landau levels. a-j**, dI/dV spectra taken between the stripes (labeled as n) subtracted by that taken on the stripe (labeled by n'). The black arrows and dashed lines mark the Landau levels.

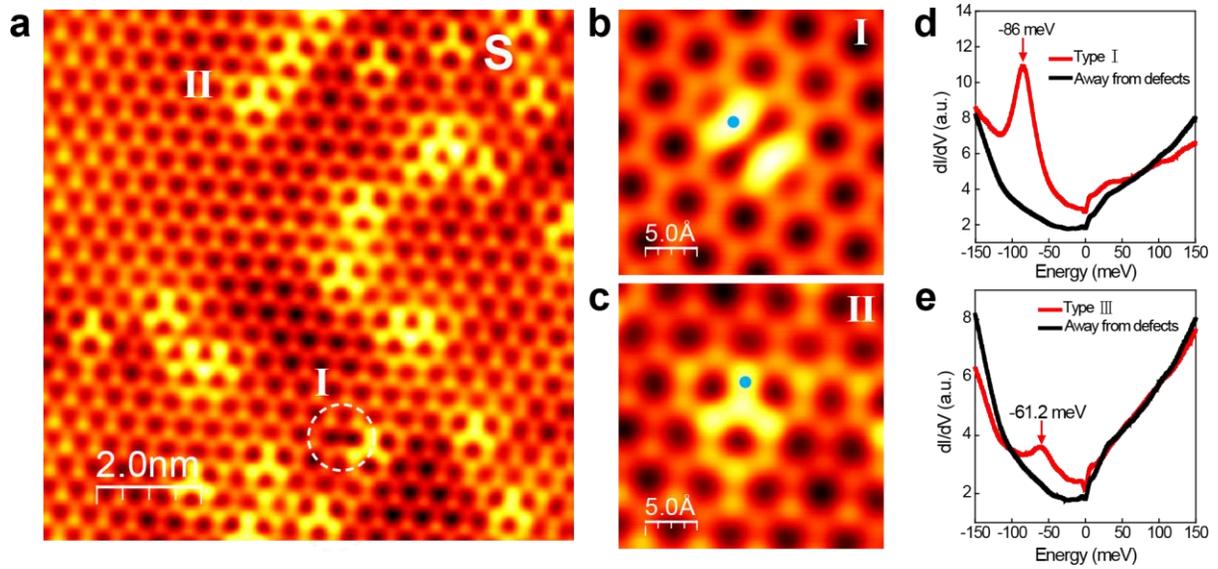

**Figure S18. Bound states associated with two types of defects on the Sn-terminated FeSn/STO(111) film.** Two types of defects commonly observed on the Sn-terminated surface. **a**, Atomic resolution image of the Sn-termination (S) with a honeycomb lattice with three-fold and two-fold symmetrical defects (dashed white circle). Setpoint: $V$ = 0.2 V, $I$ = 3.0 nA. **b**, STM imaging showing bound states of the Sn di-vacancy, setpoint: $V$ = 2.0 V, $I$ = 10 pA. **c**, STM imaging showing bound states of the substitutional defect, setpoint: $V$ = 2.0 V, $I$ = 10 pA. **d-e**, dI/dV spectra measured at the cyan dots in **b** and **c**, respectively.

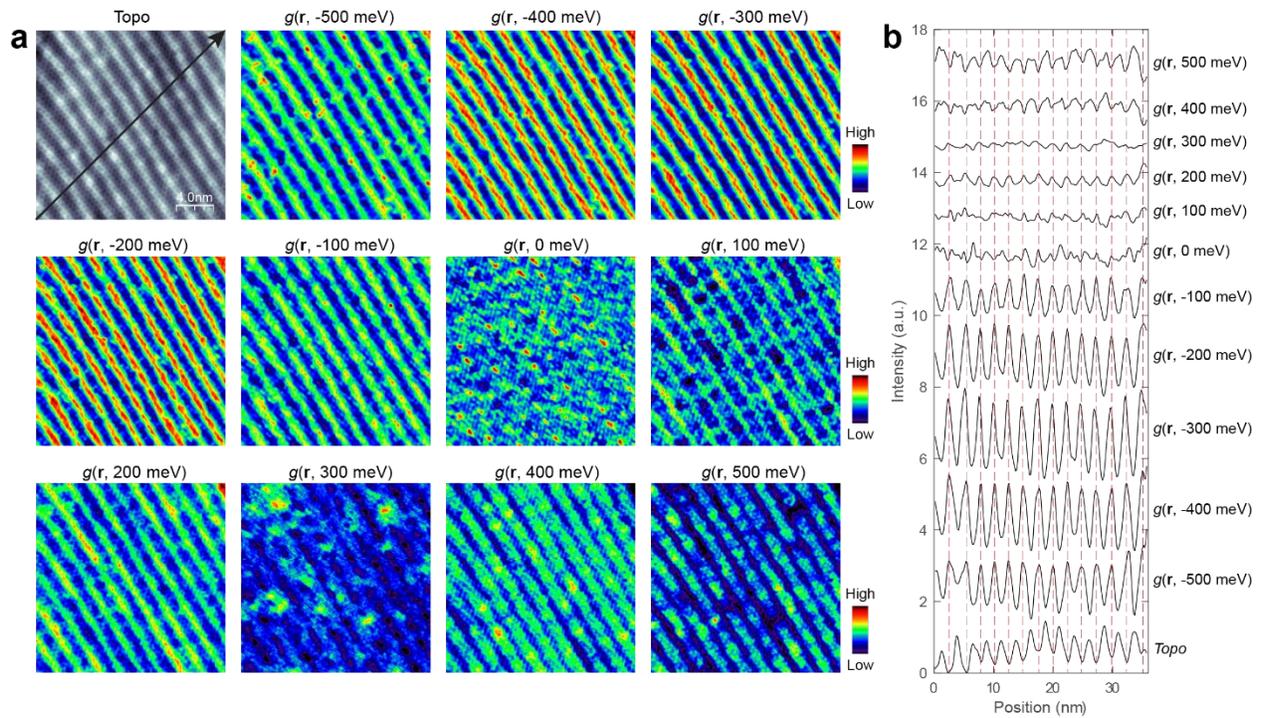

**Figure S19. Spatial-resolved dI/dV maps. a**, Topographic STM image showing periodic stripe modulations. Setpoint: $V$ = 500 mV, $I$ = 5.0 nA. **b**, Line profiles along the black arrow in (**a**) and the energy-dependent dI/dV maps. The dashed red lines mark the valley position from topography. Between the energy ranges [-500 meV, -100 meV] and [100 meV, 300 meV], the valley position in the topography corresponds to higher density of states. At Fermi level $g(\mathbf{r}, 0\text{ meV})$, the modulation is almost gone. At the energy window [400 meV, 500 meV], the valley position in the topography corresponds to lower density of states. There is a π-phase shift between the energy ranges of [400 meV, 500 meV] with the other energies. Setpoint: $V$ = 500 mV, $I$ = 5.0 nA, $V_{mod}$ = 10 mV.

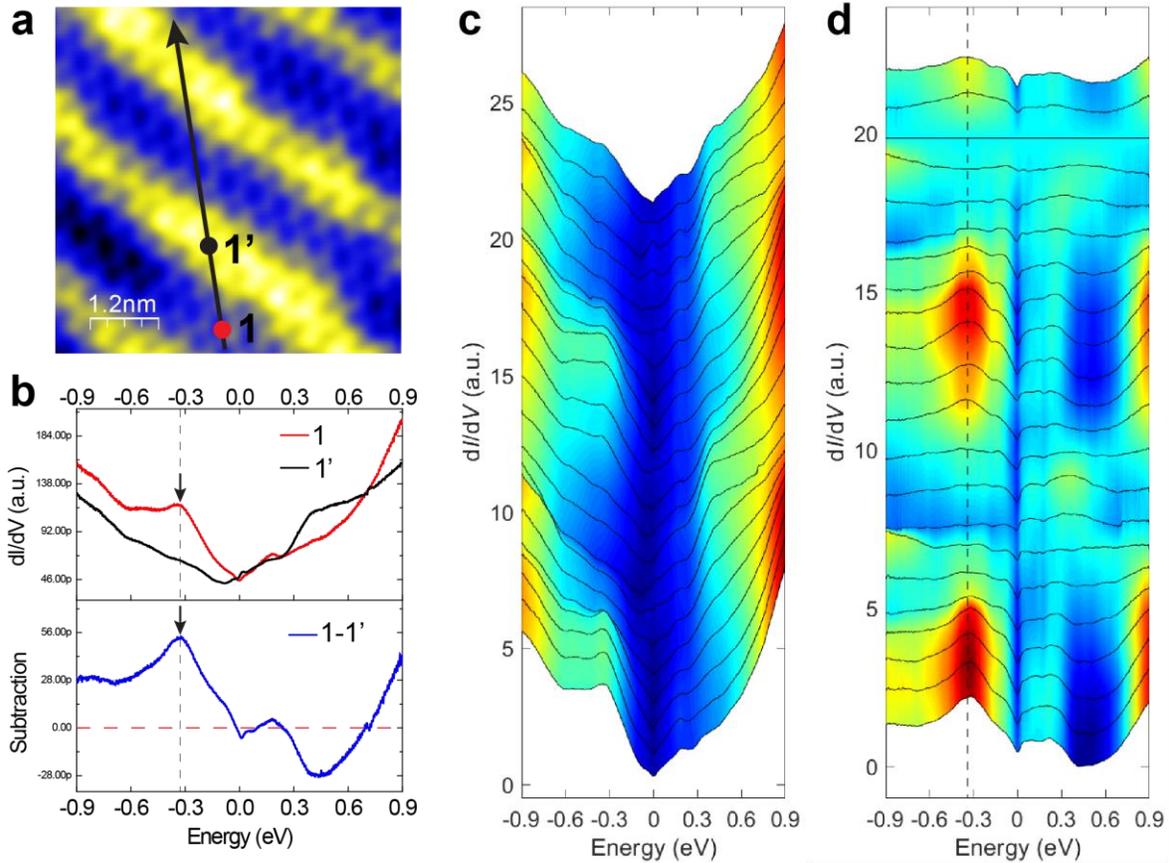

**Figure S20. Another example of Landau levels by pseudo-magnetic field. a**, Topographic STM image revealing stripes. Setpoint: $V$ = 0.5 V, $I$ = 6.0 nA. **b**, dI/dV spectra taken at the red and black dots (upper panel) and their difference spectrum (lower panel). **c**, A series of dI/dV spectra taken along the black arrow in (**a**). The peak marked by the black arrow is assigned to the $0^{th}$ landau level. **d**, The dI/dV spectra in (**c**) subtracted by the dI/dV spectrum taken on the stripe. The dashed line clearly reveals the $0^{th}$ landau level which appears between the stripes and is absent on the stripe.

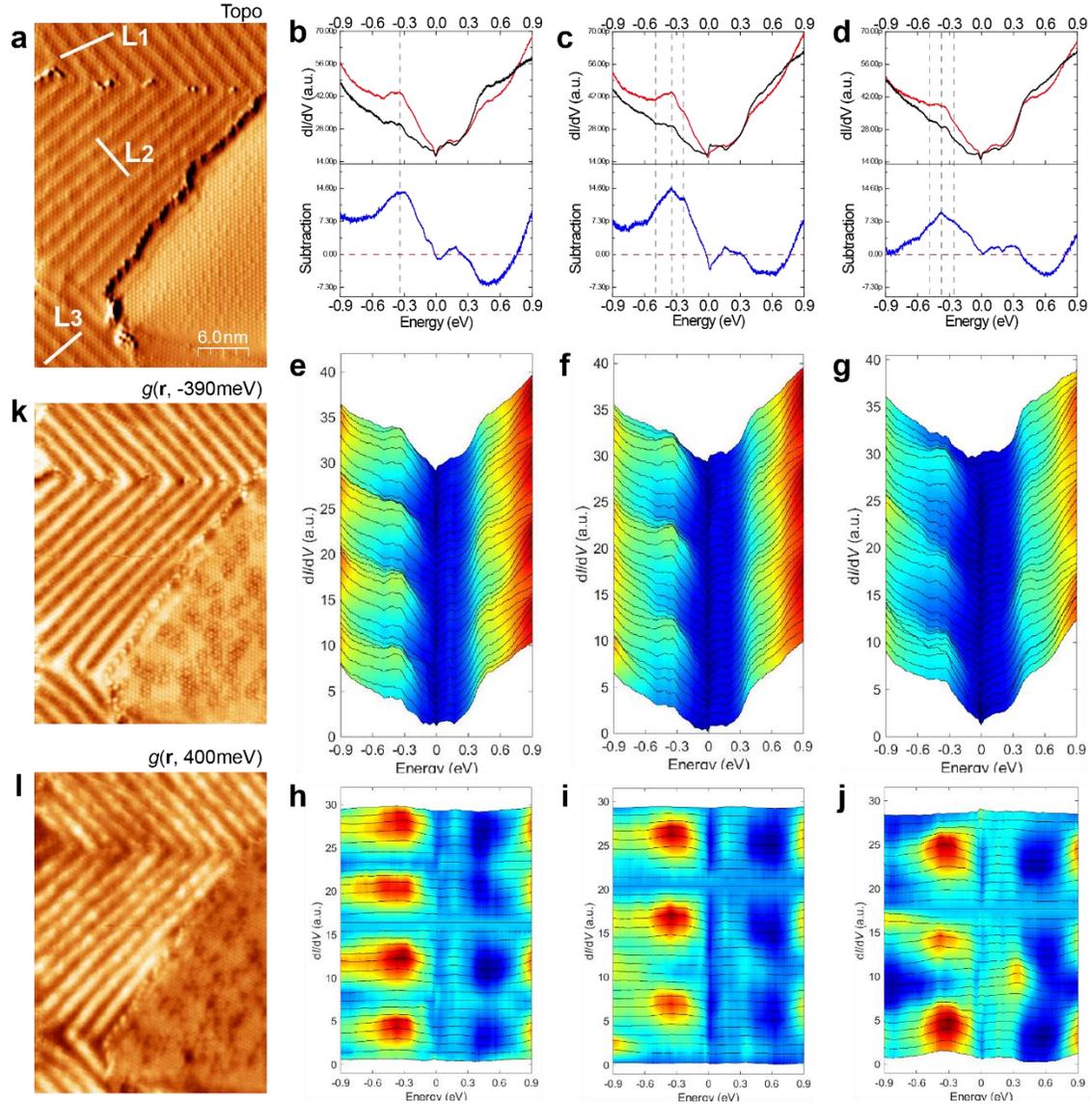

**Figure S21. Laudau levels independent on the stripe orientations. a**, Topographic STM images revealing the stripes with different orientations. Setpoint: *V* = 0.2 V, *I* = 7.0 nA. **b**-**d**, dI/dV spectra taken on the stripes (red curves) and the region between the stripes (black curve) for the line L1, L2 and L3, respectively. **e**-**g**, Line dI/dV spectra taken along the line L1, L2 and L3, respectively. **h**-**j**, A series of dI/dV spectra subtracted by the spectra taken between the stripes. **k**-**l**, dI/dV map at the energy of -390 meV and 400 meV. The contrast of the stripes reverses between the two energies.